\documentclass[aps,nofootinbib,notitlepage,superscriptaddress,floatfix,showkeys]{revtex4-1}
\usepackage{amsmath, amssymb}
\usepackage{bbm}
\usepackage{graphicx}
\usepackage{geometry}
\usepackage{xcolor}
\usepackage{hyperref}
\usepackage{cleveref}
\usepackage{orcidlink}
\def\thesection{\arabic{section}}

\renewcommand{\thetable}{\arabic{table}}%

\begin{document}
\raggedbottom
\title{Investigation of Indian stock markets using topological data analysis and geometry-inspired network measures}

\author{Saumitra Kulkarni\,\orcidlink{0000-0001-8859-3207}}
\affiliation{The Institute of Mathematical Sciences (IMSc), Chennai 600113, India}
\author{Hirdesh K. Pharasi\,\orcidlink{0000-0002-6937-9237}}
\affiliation{School of Engineering and Technology, BML Munjal University, Gurugram 122413, India}
\author{Sudharsan Vijayaraghavan\,\orcidlink{0009-0005-2331-3469}}
\affiliation{The Institute of Mathematical Sciences (IMSc), Chennai 600113, India}
\author{Sunil Kumar\,\orcidlink{0000-0001-8240-0945}}
\affiliation{Department of Physics, Ramjas College, University of Delhi, Delhi 110007, India}
\author{Anirban Chakraborti\,\orcidlink{0000-0002-6235-0204}}
\thanks{Corresponding authors: Anirban Chakraborti (\href{mailto:anirban@jnu.ac.in}{anirban@jnu.ac.in}) and Areejit Samal (\href{mailto:asamal@imsc.res.in}{asamal@imsc.res.in})}
\affiliation{School of Computational and Integrative Sciences, Jawaharlal Nehru University, New Delhi 110067, India}
\author{Areejit Samal\,\orcidlink{0000-0002-6796-9604}}
\thanks{Corresponding authors: Anirban Chakraborti (\href{mailto:anirban@jnu.ac.in}{anirban@jnu.ac.in}) and Areejit Samal (\href{mailto:asamal@imsc.res.in}{asamal@imsc.res.in})}
\affiliation{The Institute of Mathematical Sciences (IMSc), Chennai 600113, India} 
\affiliation{Homi Bhabha National Institute (HBNI), Mumbai 400094, India}

\begin{abstract}
Financial markets are a quintessential example of complex systems wherein strategic connections arising through the trading (buying and selling) of different stocks gives rise to dynamic correlation patterns in stock prices. Network science based approaches have been employed for more than two decades to investigate the stock market as an inferential network, with nodes representing stocks and edges as pairwise cross-correlations between stock prices. However, a recent focus has been the characterization of the higher-order structure, beyond pairwise correlation, of such networks. To this end, geometry-inspired measures (such as discrete Ricci curvatures) and topological data analysis (TDA) based methods (such as persistent homology) have become attractive tools for characterizing the higher-order structure of networks representing the financial systems. In this study, our goal is to perform a comparative analysis of both these approaches, especially by assessing the fragility and systemic risk in the Indian stock markets, which is known for its high volatility and risk. To achieve this goal, we analyze the time series of daily log-returns of stocks comprising the National Stock Exchange (NSE) and the Bombay Stock Exchange (BSE). Specifically, our aim is to monitor the changes in standard network measures, edge-centric discrete Ricci curvatures, and persistent homology based topological measures computed from cross-correlation matrices of stocks. In this study, the edge-centric discrete Ricci curvatures have been employed for the first time in the analysis of the Indian stock markets. The Indian stock markets are known to be less diverse in comparison to the US market, and hence provides us an interesting example. 
Our results point that, among the persistent homology based topological measures, persistent entropy is simple and more robust than $L^1$-norm and $L^2$-norm of persistence landscape. In a broader comparison between network analysis and TDA, we highlight that the network analysis is sensitive to the way of constructing the networks (threshold or minimum spanning tree), as well as the threshold values used to construct the correlation-based threshold networks. On the other hand, the persistent homology is a more robust approach and is able to capture the higher-order interactions and eliminate noisy data in financial systems, since it does not take into account a single value of threshold but rather a range of values.\\
\end{abstract}



\maketitle

\section{Introduction} \label{sec:introduction}

Econophysics is an interdisciplinary field that applies concepts and methods from physics to the study of economic and financial systems. In particular, the field aims to provide insights into the dynamics, patterns, and emergent behaviors of financial markets, income distributions, and other economic phenomena.
A prominent approach in econophysics is to treat a financial market or an economic system as a \textit{complex system} \cite{mantegna1999introduction, sinha2010econophysics, chakraborti2011econophysics}. 
A complex system is typically composed of many interacting components or agents that give rise to intricate and often unpredictable behaviors \cite{gell2002complexity, vemuri2014modeling, chakrabarti2023data}. 
Complex systems can exhibit emergent properties, where the collective behavior of the system as a whole is not directly deducible from the behavior of its individual components. 
Within the context of econophysics, the behavior of stock prices, market indices, currency exchange rates, and other financial variables have been analyzed using concepts such as tipping points, feedback loops, contagion effects, and network science \cite{battiston2016complexity}. 
Such approaches help in uncovering underlying patterns, trends, and potential instabilities within economic and financial systems.

Using the framework of network science in the context of financial systems, a stock market can be viewed as a complex network wherein the nodes represent the stocks and an edge between a pair of stocks represents the cross-correlation strength between them. Such interactions can be identified by utilizing the cross-correlations of fluctuations in the stock prices. Since these cross-correlations change over time, the underlying dynamics of the market gives rise to correlation-based networks that evolve over time. In the last two decades, there have been numerous studies on such networks, and therefore, network science has become an important and widely-used tool to model and analyze complex financial systems \cite{mantegna1999hierarchical, plerou2002random, onnela2003dynamics, tumminello2005tool, chi2010network, sandhu2016ricci, samal2021rsos, samal2021frontiers}. As the network corresponding to a cross-correlation matrix for the stocks is weighted yet completely connected (as there is at least some correlation between any pair of stocks in the market), there has been extensive research towards filtration of the relevant connections from such completely connected structures and investigation of the structural properties of resulting networks \cite{mantegna1999hierarchical, onnela2003asset, onnela2004clustering, tumminello2005tool, bonanno2003topology, lee2018state, silva2016structure, wang2018correlation}.
 
Over the last decade, geometry-inspired measures have been introduced as powerful tools to analyze the structure of real-world complex networks \cite{bianconi2015interdisciplinary, sreejith2016forman, bianconi2017emergent, boguna2021network}. 
In complex systems, an interaction can involve three or more components, and such an interaction cannot be completely captured by an edge embodying pairwise interaction among two components in the graph representation. 
Therefore, there has been significant interest towards characterizing the higher-order interactions in complex networks, including efforts to analyze real-world systems using higher-order network representations such as hypergraphs and simplicial complexes \cite{kannan2019persistent, kartun2019beyond, bianconi2021higher, battiston2020networks,
battiston2021physics}. In particular, discrete Ricci curvatures are powerful tools to characterize the higher-order interactions and relate such structure to the associated dynamics of complex networks \cite{samal2018comparative, yadav2022poset}. 
In recent times, discrete Ricci curvatures have been applied to complex networks from various domains such as gene regulatory networks implicated in cancer \cite{sandhu2015graph}, structural and functional connectivity networks of the brain \cite{farooq2019network, yadav2023discrete, elumalai2022graph} and internet topology \cite{ni2015ricci}. 
Furthermore, discrete Ricci curvatures have also been used to predict protein-ligand binding affinity \cite{wee2021ollivier, wee2021forman} and for community detection in complex networks \cite{ni2019community, sia2019ollivier, tian2023mixed}. 
Importantly, multiple studies have also assessed the ability of different notions of discrete Ricci curvatures to serve as indicators of financial market fragility and systemic risk \cite{sandhu2016ricci, samal2021rsos, samal2021frontiers}. 
For instance, Sandhu \textit{et al.} \cite{sandhu2016ricci} have studied the evolution of Ollivier-Ricci curvature \cite{ollivier2007ricci, ollivier2009ricci}  in correlation-based networks of the USA S\&P-500 stock market to show that Ollivier-Ricci curvature is an excellent indicator of market fragility. 
Subsequently, Samal \textit{et al.} \cite{samal2021rsos} have analyzed the USA S\&P-500 and the Japanese Nikkei-225 stock markets by considering three additional notions of discrete Ricci curvatures, namely Forman-Ricci \cite{forman2003bochner, sreejith2016forman}, Haantjes-Ricci \cite{haantjes1947distance, saucan2021simple} and Menger-Ricci curvatures \cite{menger1930untersuchungen, saucan2020simple, saucan2021simple} along with Ollivier-Ricci curvature. 
Samal \textit{et al.} \cite{samal2021rsos} concluded that all four discrete Ricci curvatures are able to capture the system-level features of the stock market, and Forman-Ricci curvature is the best indicator among the four discrete Ricci curvatures, of systemic volatility and fragility in both markets. In a separate study, Samal \textit{et al.} \cite{samal2021frontiers} employed the four aforementioned discrete Ricci curvatures to correlation-based networks of global financial indices and concluded that these geometry-inspired network measures can also be effectively used as indicators of fragility in global financial indices. 

Topological Data Analysis (TDA) is an interdisciplinary field that lies at the confluence of mathematics, computer science, statistics, data science and other related fields \cite{patania2017topological}. The main goal of TDA is to study the topological features or the \textit{shape} of a multidimensional dataset using ideas from algebraic topology and computational geometry. In this work, we are particularly interested in persistent homology, a widely-used method in the area of TDA. Persistent homology provides a theoretical framework and various computational tools to extract the topological features from a higher-dimensional dataset, and the method is robust to noise arising from stochastic nature of input dataset \cite{edelsbrunner2008persistent, carlsson2009topology}. Persistent homology has been successfully applied in wide range of scientific fields including biology \cite{chan2013topology,topaz2015topological,camara2016inference,crawford2016topological,xia2018multiscale}, neuroscience \cite{petri2014homological,lord2016insights,giusti2016two,bendich2016persistent,yoo2016topological}, physics \cite{kramar2013persistence,donato2016persistent,hiraoka2016hierarchical,maletic2016persistent}, computer science \cite{chung2009persistence,zhu2013persistent}, and engineering \cite{de2007coverage,bhattacharya2015persistent,pokorny2016topological}. In addition, persistent homology has been applied in the fields of economics and finance. In particular, the method has been used to identify early warning signals of market instabilites \cite{gidea2017topological, gidea2018topological, guo2020empirical, guo2021analysis, katz2021time}. Gidea \cite{gidea2017topological} developed a TDA-based method for the detection of early warning signals for critical transitions in financial data. In a subsequent paper, Gidea and Katz \cite{gidea2018topological} performed an analysis of four major US stock market indices during the crashes of 2000 and 2007-2009. Specifically, the authors used persistence landscape \cite{bubenik2015statistical} and its $L^p$-norms to quantify the temporal changes prior to a financial meltdown. Apart from the above-mentioned studies, TDA has also been employed to study market dynamics \cite{majumdar2020clustering}, corporate failures \cite{qiu2020refining}, investment strategies \cite{baitinger2021better}, and asset allocation \cite{goel2020topological}. 

In the present contribution, we examine the daily log-returns for a set of stocks comprising the two Indian stock markets namely, the National Stock Exchange (NSE) over a period of 18 years and the Bombay Stock Exchange (BSE) over a period of 23 years. 
In particular, we monitor the changes occurring in the two Indian stock markets by analysing the fluctuations in standard network measures, discrete Ricci curvatures, and persistent homology based topological measures. 
From a network geometry perspective, we use edge-centric discrete Ricci curvatures to investigate correlation-based threshold networks of Indian stock markets. 
From a TDA perspective, we use the $L^p$-norms of persistence landscape and the notion of persistent entropy \cite{rucco2016characterisation, chintakunta2015entropy} to capture the changes in cross-correlations among the stocks in the respective Indian markets. 
Notably, we compare the network geometry based analysis with persistent homology, to monitor the fragility of the two Indian markets, and moreover, as indicators of crashes occurring in stock markets.  
Such analysis could, in turn, establish a connection between the health of financial markets and the growth or decline of the Indian economy. 
Moving forward, our effort might aid in assessing how specific market crises impact the multi-scale financial and economic phenomena.


\begin{figure*}
    \centering
    \includegraphics[width=0.89\textwidth,keepaspectratio]{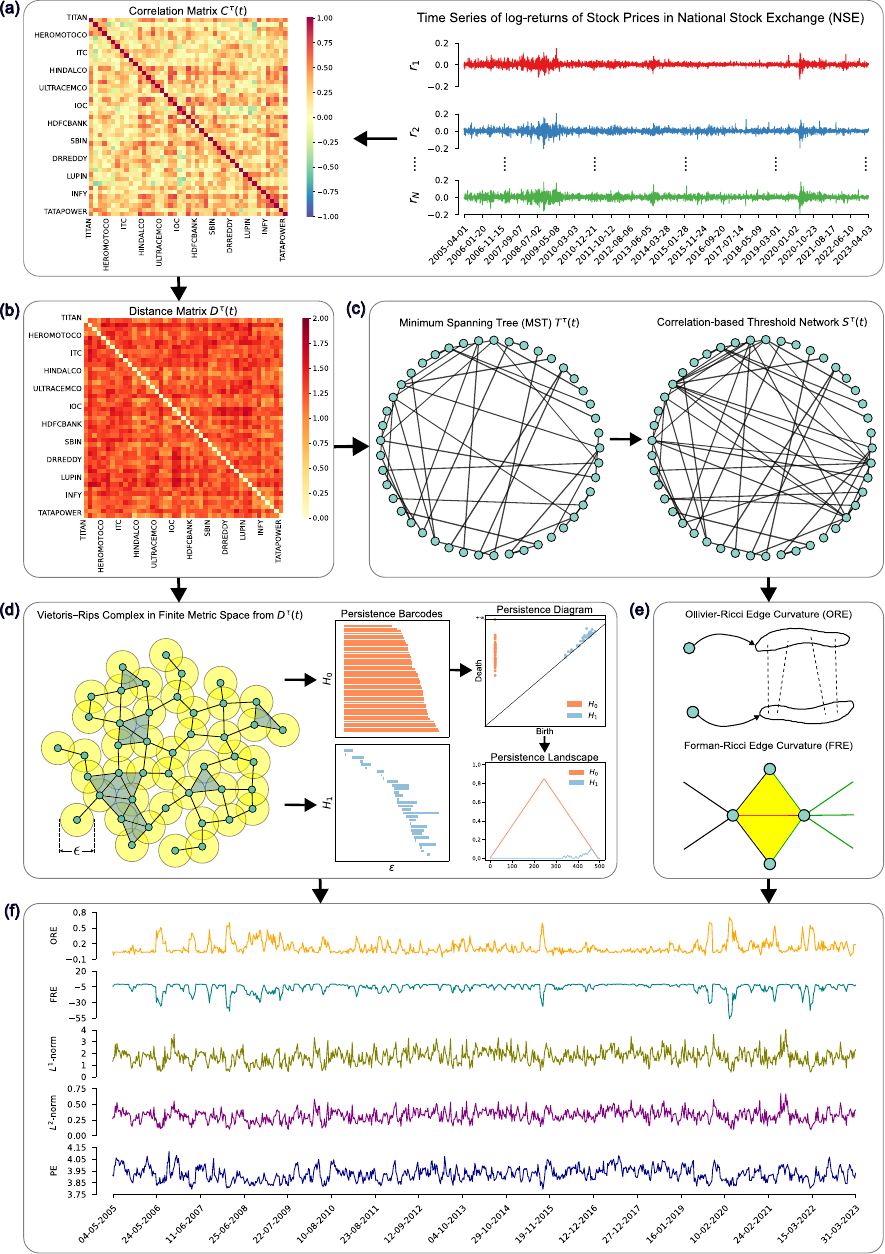}
    \caption{\textbf{Schematic overview of the computation of edge-centric discrete Ricci curvatures and persistent homology based topological measures from the time series of log-returns of stocks from the National Stock Exchange (NSE).} \textbf{(a)} Time series spanning over a 18-year period (2005–2023) and a cross-correlation matrix $C^{\tau}(t)$ for an arbitrarily chosen epoch. \textbf{(b)} Computation of distance matrix $\scriptstyle D^{\tau}(t) = \sqrt{2 (1 - C^{\tau}(t))}$. \textbf{(c)} Construction of a correlation-based threshold network $S^{\tau}(t)$ by extracting minimum spanning tree (MST) $T^{\tau}(t)$ from $D^{\tau}(t)$ and adding edges with $C^{\tau}_{ij} (t) \geq 0.65$ to it. \textbf{(d)} Computation of the persistent homology using Vietoris-Rips Complex of $D^{\tau}(t)$, the persistence barcodes for $H_0$ and $H_1$ homology groups, the persistence diagram, and the persistence landscape. \textbf{(e)} Computation of average Ollivier-Ricci edge curvature (ORE) and average Forman-Ricci edge curvature (FRE) on $S^{\tau}(t)$. \textbf{(f)} Evolution of ORE, FRE, $L^1$-norm and $L^2$-norm of persistence landscape, and persistent entropy (PE) of persistence barcodes over the time epochs of size $\tau = 22$ days and an overlapping shift of $\Delta{\tau} = 5$ days. In this figure, $C^{\tau}(t), D^{\tau}(t), T^{\tau}(t)$, and $S^{\tau}(t)$ shown in parts (a)--(c) correspond to the very first epoch in the time series.}
    \label{fig:graphical_abstract}
\end{figure*}

\section{Methods} \label{sec:methods}

\subsection{Data description}

The data used in this study was collected from the Yahoo finance database \cite{YahooFinanceDatabase} for two Indian stock markets: the National Stock Exchange (NSE) and the Bombay Stock Exchange (BSE). 
We studied the NSE market for a period of $T = 5,813$ working days spanning over 18 years starting from $1^{\text{st}}$ April 2005 to $3^{\text{rd}}$ April 2023, which includes the daily closing prices of $N=45$ stocks (see SI Table \labelcref{tab:NSE_Stocks}). In case of BSE market, we analysed the daily adjusted closing prices of $N=77$ stocks for a period of $T = 4,444$ working days spanning over 23 years starting from $4^{\text{th}}$ January 2000 to $6^{\text{th}}$ April 2023 (see SI Table \labelcref{tab:BSE_Stocks_1}). In both markets, we considered only the stocks that were traded for the corresponding entire time span. Further, any stock with a consecutive stationary value (or fixed value) for more than two months (i.e., $44$ working days) and with high non-stationary behavior (price return  $> 60$\%) was removed from the BSE dataset. Housing Development Finance Corporation Limited (HDFC) in NSE was the only stock whose price remained constant for $499$ consecutive days. Additionally, for the index log-return ($r$) in our analysis, we used the NIFTY 50 index, a weighted average of the 50 largest Indian companies listed on the NSE, and the SENSEX index, a weighted average of the 30 largest Indian companies listed on the BSE.


\subsection{Cross-correlation and distance matrices}

We constructed a time series of logarithmic returns, or simply log-returns, as $r_i(t) = \ln g_i(t) - \ln g_i (t-1)$, wherein $g_i(t)$ is the daily closing price of stock $i = 1, 2, \ldots, N$ on day $t$, with $N = 45$ stocks from NSE and $N=77$ stocks from BSE.
Thereafter, we constructed the cross-correlation matrices using equal-time Pearson cross-correlation coefficient given by:
\begin{equation}
    \label{eq:pearson_cross_corr}
    C_{ij}^{\tau}(t) = \frac{\langle r_i r_j \rangle - \langle r_i \rangle \langle r_j \rangle}{\sigma_i \sigma_j}
\end{equation}
where both $i, j = 1, 2, \ldots, N$. 
The averages $\langle \cdot \rangle$ and the standard deviations $\sigma_i$, $\sigma_j$ are computed over a period of $\tau = 22$ days with end date denoted by $t$. 
Furthermore, we constructed the `ultrametric' distance matrix using the distance measure given by Eq.(\ref{eq:distance_measure}) whose values range from 0 to 2.
\begin{equation}
    \label{eq:distance_measure}
    D_{ij}^{\tau}(t) = \sqrt{2 (1 - C_{ij}^{\tau}(t))}.
\end{equation}
We generated the correlation matrices for overlapping windows of $\tau = 22$ days with a rolling shift of $\Delta \tau = 5$ days (see Figure \ref{fig:graphical_abstract}).
Thus, in case of the NSE market for the period of 18 years, we obtained 885 distance matrices, and in case of the BSE market for the period of 23 years, we obtained 1159 distance matrices. 
We constructed the correlation-based threshold networks of stocks using $C^{\tau}(t)$ and $D^{\tau}(t)$ matrices, and thereafter, evaluated the two discrete Ricci curvatures and several standard network measures (see Figure \ref{fig:graphical_abstract}).
Furthermore, we used these distance matrices $D^{\tau}(t)$ to compute the persistent homology, and thereafter, evaluated the three topological measures (see Figure \ref{fig:graphical_abstract}). 


\subsection{Construction of correlation-based threshold networks}

For a given time epoch of $\tau$ days ending on trading day $t$, one can view the corresponding symmetric distance matrix $D^{\tau}(t)$ as an undirected complete graph $G^{\tau}(t)$ of $N$ stocks. 
The weight of an edge between any pair of stocks $i$ and $j$ is determined by the distance $D_{ij}^{\tau}(t)$ given in Eq.(\ref{eq:distance_measure}), and the Minimum Spanning Tree (MST) $T^{\tau}(t)$ can be extracted from the resulting graph by applying Prim's algorithm \cite{prim1957shortest}. 
However, the MST is a tree consisting of a subset of $N-1$ edges from $G^{\tau}(t)$ connecting the $N$ nodes such that the sum of the edge weights for the $N-1$ edges is the minimum possible. 
Since MST is just a tree with $N-1$ edges devoid of cycles, it may fail to capture all significant edges in the graph $G^{\tau}(t)$. 
To address this issue, we augment the edges in the MST $T^{\tau}(t)$, by adding edges within $G^{\tau}(t)$ with $C_{ij}^{\tau}(t) \ge 0.65$, to create a correlation-based threshold network $S^{\tau}(t)$. 
Following this procedure, we generated 885 networks for the NSE market over the 18-year period and 1159 networks for the BSE market over the 23-year period, with overlapping windows of $\tau = 22$ days and a rolling shift of $\Delta \tau = 5$ days (see Figure \ref{fig:graphical_abstract}).
These networks were utilized to compute the two discrete Ricci curvatures and several standard network measures. 


\subsection{Traditional market indicators, standard network measures and discrete Ricci curvatures}

We computed three traditional market indicators from the correlation matrices $C^{\tau}(t)$ for both NSE and BSE markets. Firstly, we computed the mean correlation ($\mu$) given by the average of the correlations in the matrix $C^{\tau}(t)$. Secondly, we computed the volatility of the market index $r$ estimated using GARCH(1,1) process. 
Thirdly, we computed the risk corresponding to the Markowitz portfolio of the stocks ($\sigma_p$), which is a proxy for the fragility or systemic risk associated with the correlation-based threshold network of stocks.
A description of GARCH$(p,q)$ process and Markowitz portfolio optimization is provided in Appendix \ref{appendix:market_indicators}.

For the correlation-based threshold networks corresponding to the NSE and BSE markets, we have computed several standard network measures such as the number of edges, edge density, average degree, average weighted degree, clique number, diameter, average clustering coefficient, modularity, communication efficiency and network entropy. All of the standard network measures computed in this study are described in Appendix \ref{appendix:network_measures}.

Notably, we have also computed two edge-centric discrete Ricci curvatures namely, Ollivier-Ricci edge curvature and Forman-Ricci edge curvature, for the correlation-based threshold networks corresponding to the NSE and BSE markets. A detailed description of these discrete Ricci curvatures is contained in Appendix \ref{appendix:discrete_ricci_curvatures}.


\subsection{Persistent homology and topological measures}

For analyzing the time series of log-returns of stocks from NSE and BSE using persistent homology, we consider set of all stocks from each market as a point cloud $S$ in a finite metric space with distance function defined by the distance matrix $D^{\tau}(t)$ for an epoch of size $\tau = 22$ days and an overlapping shift of $\Delta{\tau} = 5$ days. We are interested in understanding the topological features of this space using persistent homology, where it associates a homology group $H_i(S)$ to the space $S$, where $i = \lbrace 0, 1, 2, \dots \rbrace$. The $H_0(S)$ group refers to the number of connected components in $S$, the $H_1(S)$ group describes the number of loops, and the $H_2(S)$ group is a count of the number of voids, and so on \cite{camara2016topological}. In this work, we focus on only two homology groups, $H_0(S)$ and $H_1(S)$. We use Vietoris-Rips complex, or simply, Rips complex \cite{vietoris1927hoheren, edelsbrunner2022computational} to extract a continuous shape from the point cloud of stocks. We examine the topological features, such as connected components and loops, that persist in the Rips complex by varying the filtration parameter $\epsilon$ defined by the unique distance values from each matrix $D^{\tau}(t)$ for an epoch of size $\tau = 22$ days and an overlapping shift of $\Delta{\tau} = 5$ days for both markets.

We use two primary tools to visualize the persistent homology of our filtration, namely the persistence barcode \cite{ghrist2008barcodes} and persistence diagram \cite{edelsbrunner2022computational}. Although both of these tools serve as valuable visualization techniques for capturing the persistence of topological features, they lack suitability for statistical analysis. As a result, we compute three topological measures, from each of these representations for distance matrix $D^{\tau}(t)$, each corresponding to epoch size $\tau = 22$ days and an overlapping shift of $\Delta{\tau} = 5$ days. Specifically, we compute the persistent entropy of the persistence barcode \cite{rucco2016characterisation, chintakunta2015entropy}, and we take $L^1$-norm and $L^2$-norm of the persistence landscape extracted from persistence diagram \cite{bubenik2015statistical}. A detailed discussion on persistent homology is provided in the Appendix \ref{appendix:persistent_homology}.

\begin{figure*}
    \centering %
    \includegraphics[width=\textwidth,keepaspectratio]{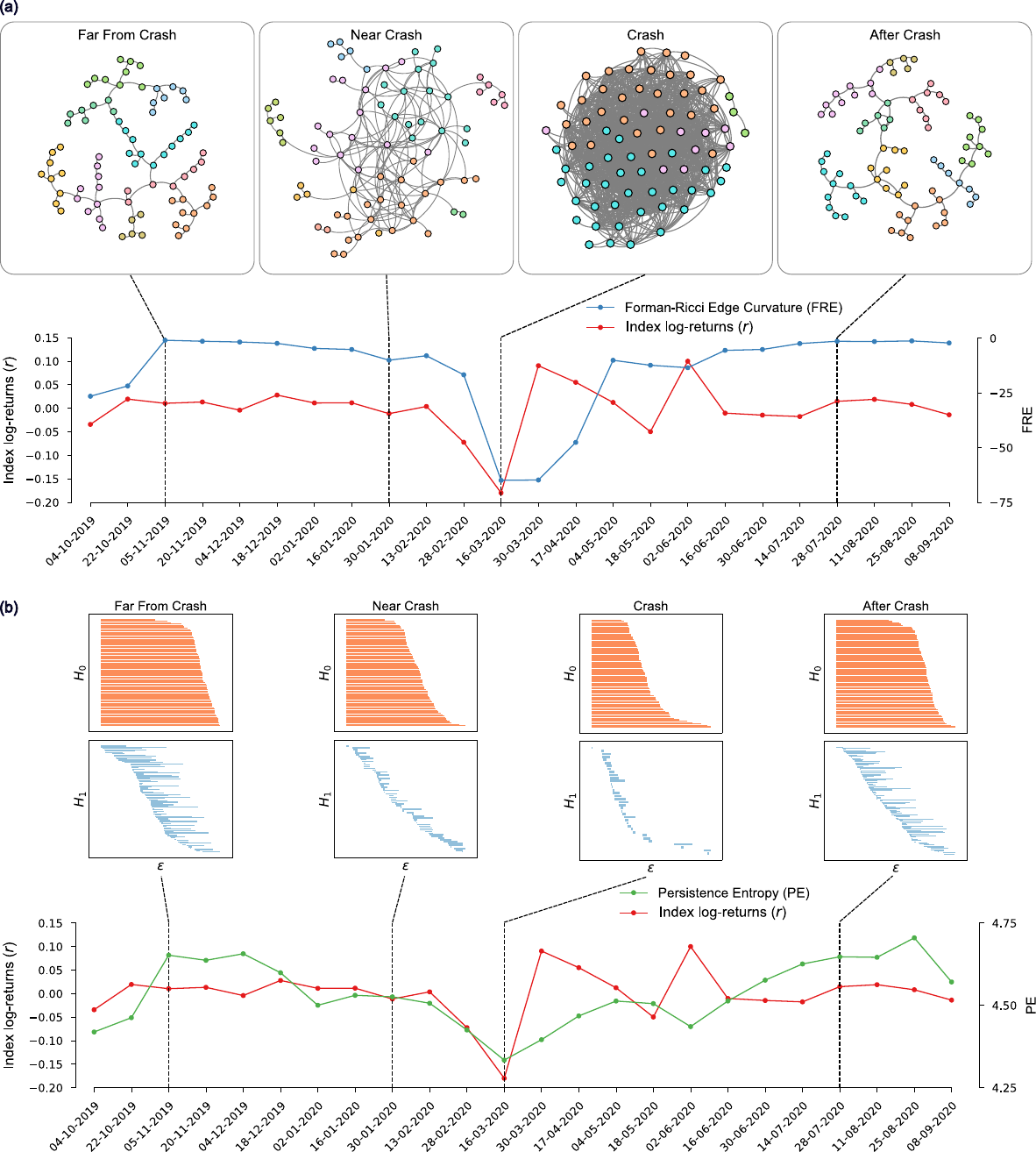}
    \caption{\textbf{Evolution of network structure and homology of correlations between stocks of Bombay Stock Exchange (BSE) through the COVID-19 Recession period at four distinct epochs of $\mathbf{\tau=22}$ days and an overlapping shift of $\mathbf{\Delta{\tau} = 5}$ days ending on trading days \textit{t}: November $5^{\text{th}}$, 2019, January $30^{\text{th}}$, 2020, March $16^{\text{th}}$, 2020, and July $28^{\text{th}}$, 2020.} \textbf{(a)} (Top panel) Visualization of correlation-based threshold networks $S^{\tau}(t)$ of stocks, constructed by adding edges with correlation $C^{\tau}_{ij}(t) \geq 0.65$ to the MST. Here, the color of the nodes correspond to the different communities detected by Louvain method \cite{blondel2008fast} for community detection. In $S^{\tau}(t)$, the edge density increases and number of detected communities decreases during a crash. (Bottom panel) Evolution of index log-returns ($r$) (red color line) and average Forman-Ricci edge curvature (FRE) (blue color line). The FRE measure, computed from $S^{\tau}(t)$, is sensitive to both local (sectoral) and global fluctuations of the market, and shows a local minimum (more negative) during the crash. \textbf{(b)} (Top panel) Visualization of persistence barcodes computed from the persistent homology of distance matrix $\scriptstyle D^{\tau}(t) = \sqrt{2 (1 - C^{\tau}(t))}$. Here, the barcodes represent birth and death of holes during the filtration for $H_0$ and $H_1$ homology groups. The number of holes reduce and persistence gets weakened with widths of the bars becoming smaller during the crash. (Bottom panel) Evolution of index log-returns ($r$) (red color line) and persistent entropy (PE) of the persistence barcodes (green color line). During the crash, $r$ has high fluctuations (high volatility) and PE also decreases significantly.}
    \label{fig:evolution_networks_barcodes}
\end{figure*}
\begin{figure*}
    \centering%
    \includegraphics[width=\textwidth,keepaspectratio]{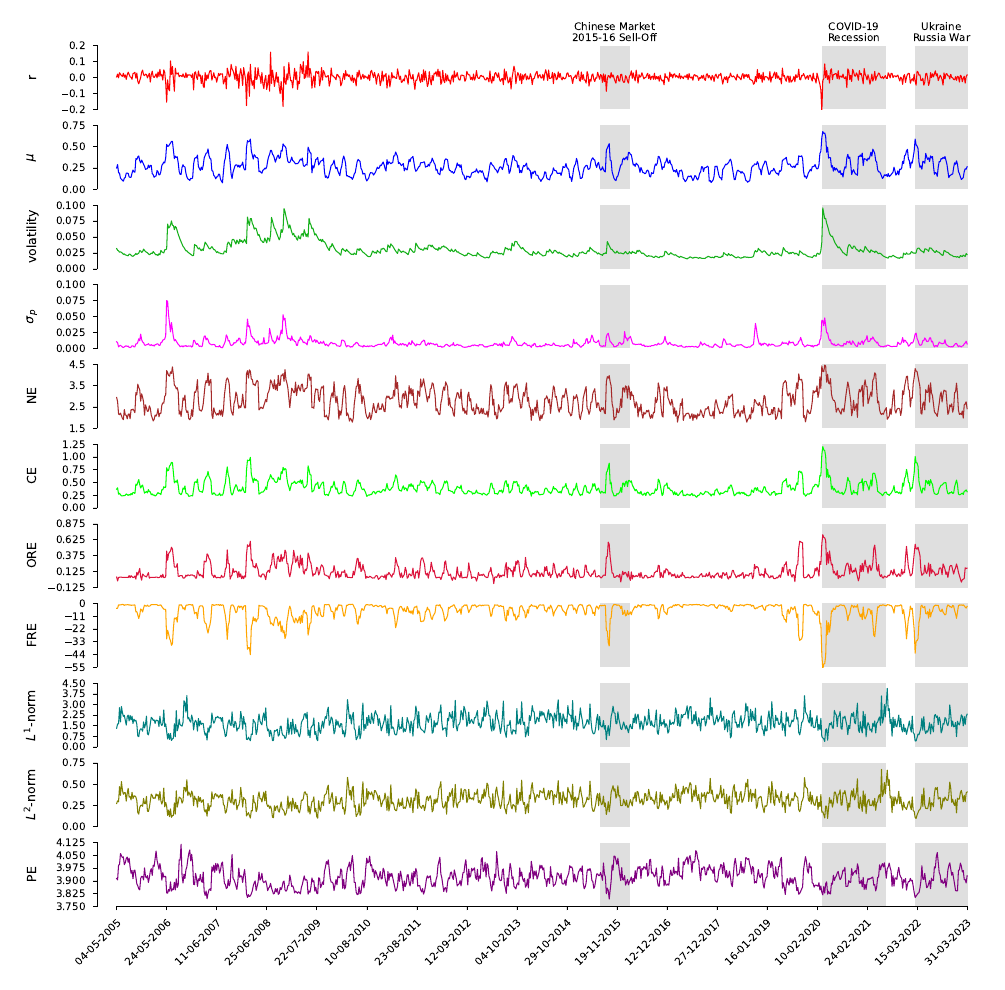}
    \caption{\textbf{Evolution of traditional market indicators, standard network measures, edge-centric discrete Ricci curvatures, and persistent homology based topological measures for the National Stock Exchange (NSE).} We evaluate standard network measures and edge-centric discrete Ricci curvatures on correlation-based threshold networks $S^{\tau}(t)$ of stocks from NSE, constructed by adding edges with correlation $C^{\tau}_{ij}(t) \geq 0.65$ to the MST. We compute topological measures from the persistent homology of the distance matrices $\scriptstyle D^{\tau}(t) = \sqrt{2 (1 - C^{\tau}(t))}$ corresponding to the epochs of size $\tau = 22$ days and an overlapping shift of $\Delta{\tau} = 5$ days. From top to bottom, we compare the index log-returns ($r$), mean market correlation ($\mu$), volatility of the market index $r$ estimated using GARCH(1,1) process, risk ($\sigma_p$) corresponding to the minimum risk Markowitz portfolio of all the stocks in the market, network entropy (NE), communication efficiency (CE), average Ollivier-Ricci edge curvature (ORE), average Forman-Ricci edge curvature (FRE), $L^1$-norm and $L^2$-norm of persistence landscape and persistent entropy (PE). The grey-shaded regions indicate the epochs of important events in the NSE market: Chinese market 2015-16 sell-off, COVID-19 Recession and Ukraine Russia war.}
    \label{fig:01_nse_time_series}
\end{figure*}
\begin{figure*}
    \centering%
    \includegraphics[width=\textwidth,keepaspectratio]{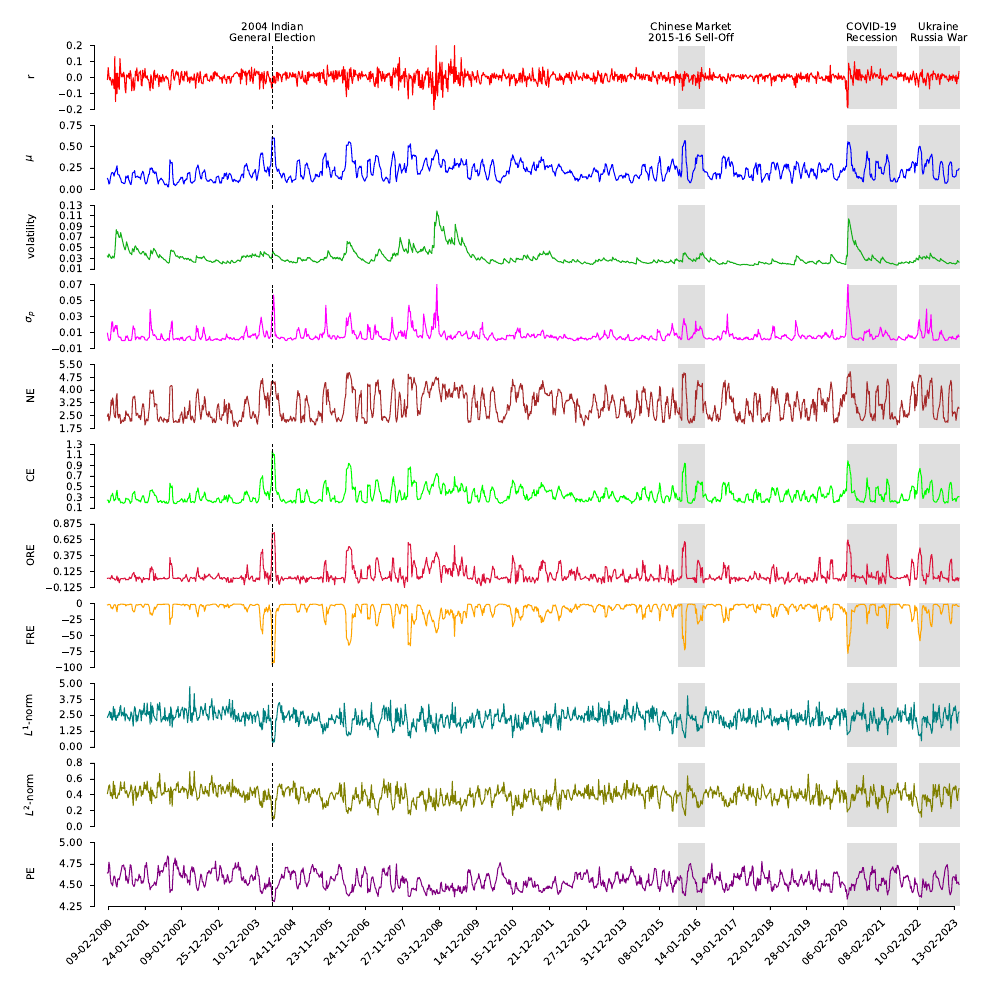}
    \caption{\textbf{Evolution of traditional market indicators, standard network measures, edge-centric discrete Ricci curvatures, and persistent homology based topological measures for the Bombay Stock Exchange (BSE).} We evaluate standard network measures and edge-centric discrete Ricci curvatures on the correlation-based threshold networks $S^{\tau}(t)$ of stocks from BSE, constructed by adding edges with correlation $C^{\tau}_{ij}(t) \geq 0.65$ to the MST. We compute topological measures from the persistent homology of the distance matrices $\scriptstyle D^{\tau}(t) = \sqrt{2 (1 - C^{\tau}(t))}$ corresponding to the epochs of size $\tau = 22$ days and an overlapping shift of $\Delta{\tau} = 5$ days. From top to bottom, we compare the index log-returns ($r$), mean market correlation ($\mu$), volatility of the market index $r$ estimated using GARCH(1,1) process, risk ($\sigma_p$) corresponding to the minimum risk Markowitz portfolio of all the stocks in the market, network entropy (NE), communication efficiency (CE), average Ollivier-Ricci edge curvature (ORE), average Forman-Ricci edge curvature (FRE), $L^1$-norm and $L^2$-norm of persistence landscape and persistent entropy (PE). The vertical dashed lines and grey-shaded regions indicate the epochs of important events in the BSE market: 2004 Indian general election, Chinese market 2015-16 sell-off, COVID-19 Recession and Ukraine Russia war.}
    \label{fig:02_bse_time_series}
\end{figure*}
\section{Results and Discussion} \label{sec:results_discussion}

In this work, we constructed two separate time series of log-returns of: (a) 45 stocks from the National Stock Exchange (NSE) over a period of 18 years (2005-2023), and (b) 77 stocks from Bombay Stock Exchange (BSE) over a period of 23 years (2000-2023). 
Next we constructed the Pearson cross-correlation matrices $C^{\tau}(t)$ for the epochs of size $\tau = 22$ days and an overlapping shift of $\Delta{\tau} = 5$ days and ending on trading days $t$. 
Using $C^{\tau}(t)$, we studied the evolution of correlation-based threshold networks $S^{\tau}(t)$ by employing network measures including edge-centric discrete Ricci curvatures, and that of the distance matrices $\scriptstyle D^{\tau}(t) = \sqrt{2 (1 - C^{\tau}(t))}$ by employing topological measures computed from persistent homology of $D^{\tau}(t)$. 
In Figure \ref{fig:graphical_abstract}, we present a schematic overview of the evaluation procedure for two edge-centric discrete Ricci curvatures from the correlation-based threshold networks $S^{\tau}(t)$ and three topological measures from the persistent homology of distance matrices $D^{\tau}(t)$ for time series of daily log-returns of stocks from the NSE market.
The time series of daily log-returns of $N$ stocks of NSE market for a 18-year period (2005-2023) and conversion of the time series of the very first epoch of $\tau = 22$ working days into a correlation matrix $C^{\tau}(t)$ using Pearson cross-correlation coefficient are shown in Figure \ref{fig:graphical_abstract}(a). 
Extraction of distance matrix $\scriptstyle D^{\tau}(t) = \sqrt{2 (1 - C^{\tau}(t))}$ corresponding to the correlation matrix $C^{\tau}(t)$ of the first epoch of $\tau = 22$ working days is shown in Figure \ref{fig:graphical_abstract}(b). 
The construction of a correlation-based threshold network $S^{\tau}(t)$, by extracting the minimum spanning tree (MST) $T^{\tau}(t)$ from the distance matrix $D^{\tau}(t)$ and subsequently adding edges to $T^{\tau}(t)$ with $C_{ij}^{\tau}(t) \geq 0.65$, is shown in Figure \ref{fig:graphical_abstract}(c). 
The computation of persistent homology by building Rips complex from the distance matrix $D^{\tau}(t)$, and computation of corresponding topological measures from persistence barcodes, persistence diagram, and persistence landascape is shown in Figure \ref{fig:graphical_abstract}(d). 
The two edge-centric discrete Ricci curvatures, namely, Ollivier-Ricci edge curvature (ORE) and Forman-Ricci edge curvature (FRE) are computed using the correlation-based threshold networks as shown in Figure \ref{fig:graphical_abstract}(e). 
Finally, in Figure \ref{fig:graphical_abstract}(f), evolution of two edge-centric discrete Ricci curvatures and persistent homology based topological measures such as $L^1$-norm and $L^2$-norm of persistence landscape, and persistent entropy (PE) of persistence barcodes are shown.

In previous studies employing edge-centric discrete Ricci curvatures to analyze stock market data, some of us have shown that both FRE and ORE are highly correlated to the traditional market indicators and are better indicators of fragility and systemic risk in the financial systems in comparison to the standard graph-theoretic measures \cite{samal2021rsos, samal2021frontiers}. 
Notably, the FRE is a simpler measure to compute as it depends only on the information of immediate neighbors of an edge under consideration, whereas the ORE is much more expensive to compute in large-scale networks \cite{samal2018comparative}. 
Amongst the persistent homology based topological measures, previously researchers have employed the $L^p$-norms of persistence landscape to quantify the temporal changes in the time series of stock market data \cite{goel2020topological, majumdar2020clustering, gidea2018topological, guo2021analysis}. 
Here, we employ an additional measure called the PE of persistence barcodes, and moreover, we argue that the PE is a better measure for capturing the higher-order interactions among the stocks of a financial market than the persistence landscape and its $L^p$-norms. 
Therefore, we start by demonstrating the ability of FRE and investigating the ability of PE, to capture the evolution of correlations between stocks of an Indian stock market.

Figure \ref{fig:evolution_networks_barcodes} shows the evolution of index log-returns with respect to: (a) average FRE of the correlation-based threshold networks and (b) PE of persistence barcodes, for the COVID-19 Recession period for the BSE market. 
The top panel in Figure \ref{fig:evolution_networks_barcodes}(a) shows the evolution in the structure of correlation-based threshold networks for the COVID-19 Recession period at four distinct epochs of size $\tau = 22$ days ending on following trading days: $5^{\text{th}}$ November 2019, $30^{\text{th}}$ January 2020, $16^{\text{th}}$ March 2020, $28^{\text{th}}$ July 2020, with threshold $C^{\tau}_{ij}(t) \geq 0.65$. 
The graph densities \cite{coleman1983estimation} and number of communities \cite{porter2009communities} in these networks are 2.6\%, 6\%, 40.8\%, 2.7\% and 8, 9, 4, 9, respectively. 
The node colors in these networks correspond to the different communities detected by Louvain method \cite{blondel2008fast} for community detection. 
The correlation-based threshold networks exhibit extremely high graph densities and low number of communities during the crash. 
The plots of index log-returns $r$ of the SENSEX index (red line) and average FRE (blue line) are shown around the COVID-19 Recession period in the bottom panel of Figure \ref{fig:evolution_networks_barcodes}. 
During a market crash, index log-returns $r$ displays high volatility and FRE shows a significant decrease (local minimum). 
Figure \ref{fig:evolution_networks_barcodes}(b) shows the evolution in homology of stock market data around the same period of COVID-19 Recession at the same distinct epochs of $\tau = 22$ days ending on following trading days: $5^{\text{th}}$ November 2019, $30^{\text{th}}$ January 2020, $16^{\text{th}}$ March 2020, $28^{\text{th}}$ July 2020, with threshold $C_{ij}^{\tau}(t) \geq 0.65$. 
The top panel of Figure \ref{fig:evolution_networks_barcodes}(b) shows the evolution in the persistence barcodes at each of these trading days. 
During the crash, the persistence barcode shows a greater variance in length of each bar and the lowest number of $p$-holes (loops) for $H_0$ and $H_1$ homology groups, respectively. 
This is because, the market becomes extremely correlated during a crash, increasing $C_{ij}^{\tau}(t)$ values in correlation matrices \cite{pharasi2018identifying}. 
As a consequence, the corresponding ultrametric distances $D_{ij}^{\tau}(t)$ in the distance matrices among the stocks also decrease. 
In the bottom panel of Figure \ref{fig:evolution_networks_barcodes}(b), we plot the index log-returns $r$ of the NIFTY50 index (red line) and the PE (green line) around the COVID-19 Recession period. 
As before, $r$ shows a volatile behavior during the crash, and consequently, the PE also decreases to a local minima, since variation in the length of the bars of persistence barcode also increases drastically.

We show evolution of traditional market indicators, standard network measures, two edge-centric discrete Ricci curvatures, and persistent homology based topological measures, namely $L^1$-norm and $L^2$-norm of persistence landscape and PE, for the NSE and BSE markets in Figures \labelcref{fig:01_nse_time_series,fig:02_bse_time_series}, respectively. 
Both figures highlight three periods of financial instability, namely Chinese market 2015-2016 sell-off, COVID-19 Recession, and Ukraine Russia War. 
Additionally, Figure \ref{fig:02_bse_time_series} highlights the financial crash of 2004 Indian general election. 
We compute standard network measures from the correlation-based threshold networks $S^{\tau}(t)$ constructed using MST and adding edges with $C^{\tau}_{ij}(t) \geq 0.65$, and topological measures from the persistent homology of the distance matrices $D^{\tau}(t)$, both for the epoch size of $\tau = 22$ days and an overlapping shift of $\Delta{\tau} = 5$ for both markets. 
In both Figures \labelcref{fig:01_nse_time_series,fig:02_bse_time_series}, we plot index log-returns ($r$), mean market correlation ($\mu$), volatility of the market index $r$ estimated using GARCH(1,1) process, risk ($\sigma_p$) corresponding to the minimum risk Markowitz portfolio of all the stocks in the respective market, network entropy (NE), communication efficiency (CE), average ORE, average FRE, $L^1$-norm and $L^2$-norm of persistence landscape and PE, from top to bottom. 
We see that $\mu$, volatility, and $\sigma_p$ increase during the periods of crisis. 
In case of standard network measures, we observe an increase in NE, CE and average ORE, whereas we see a decrease in the average FRE during crisis period in comparison to normal periods. 
Moving to the persistent homology based topological measures, both $L^1$-norm and $L^2$-norm of persistence landscape and PE decrease during the crisis period. 
Earlier studies \cite{sandhu2016ricci, samal2021rsos, samal2021frontiers} have shown that ORE and FRE, along with multiple standard network measures, are good indicators of market instability. 
Here, we have also studied the temporal evolution of the persistent homology based topological measures of distance matrices $D^{\tau}(t)$ of stocks, and found that $L^1$-norm and $L^2$-norm of persistence landscape and PE are also excellent indicators of market instability. 
In SI Figures \labelcref{fig:01_nse_time_series_075_SI,fig:01_nse_time_series_085_SI,fig:02_bse_time_series_075_SI,fig:02_bse_time_series_085_SI}, we report the results for similar correlation-based threshold networks with $C^{\tau}_{ij}(t) \geq 0.75$ and $C^{\tau}_{ij}(t) \geq 0.85$. 
In SI Figures \labelcref{fig:04_nse_gt_time_series_065_SI,fig:04_nse_gt_time_series_075_SI,fig:04_nse_gt_time_series_085_SI,fig:05_bse_gt_time_series_065_SI,fig:05_bse_gt_time_series_075_SI,fig:05_bse_gt_time_series_085_SI}, we also show that a few other standard network measures, such as number of edges, edge density, average degree, average weighted degree, clique number, network diameter, average clustering coefficient, and modularity are also good indicators of market instability for both NSE and BSE markets.

\begin{figure*}
    \centering
    \includegraphics{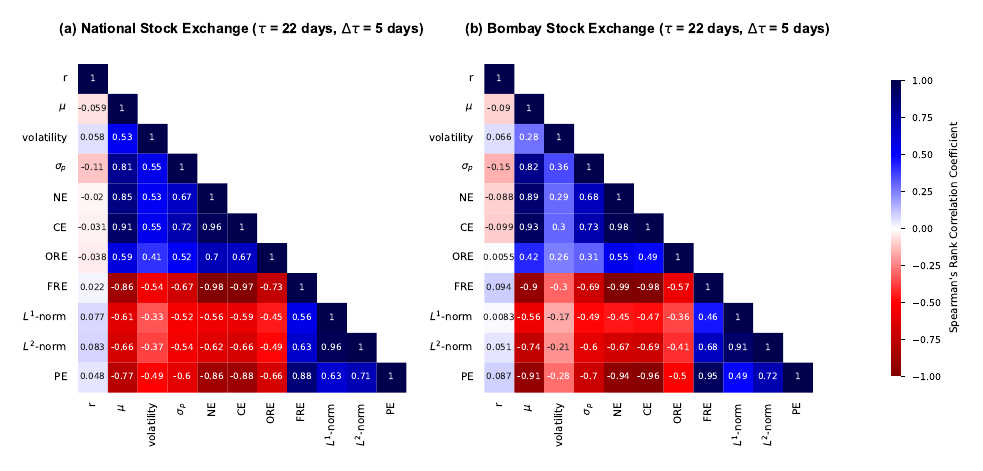}
    \caption{\textbf{Correlograms of (a) National Stock Exchange (NSE) and (b) Bombay Stock Exchange (BSE).} We show values of Spearman's rank correlation coefficient between all possible pairs of the traditional market indicators (index returns $r$, mean market correlation $\mu$, volatility, and minimum risk portfolio $\sigma_p$), standard network measures (network entropy NE, communication efficiency CE), edge-centric discrete Ricci curvatures (ORE, FRE) and persistent homology based topological measures ($L^1$-norm, $L^2$-norm, PE). The standard network measures and edge-centric discrete Ricci curvatures are computed on the correlation-based threshold networks $S^{\tau}(t)$ of stocks, constructed by adding edges with correlation $C^{\tau}_{ij}(t) \geq 0.65$ to the MST. The topological measures are computed from the persistent homology of distance matrices $\scriptstyle D^{\tau}(t) = \sqrt{2 (1 - C^{\tau}(t))}$. Each $S^{\tau}(t)$ and $D^{\tau}(t)$ correspond to the epoch size of $\tau = 22$ days with overlapping shift $\Delta{\tau} = 5$ days.}
    \label{fig:03_correlograms}
\end{figure*}

Figure \ref{fig:03_correlograms} show correlograms for traditional market indicators (index log-returns $r$, mean market correlation $\mu$, volatility, minimum risk $\sigma_p$), standard network measures (NE and CE), edge-centric discrete Ricci curvatures (ORE and FRE), and persistent homology based topological measures ($L^1$-norm, $L^2$-norm and PE), computed for epoch size of $\tau = 22$ days and an overlapping shift of $\Delta{\tau} = 5$ days for the NSE and BSE markets. 
Here, the standard network measures and both the edge-centric discrete Ricci curvatures are computed for correlation-based threshold networks $S^{\tau}(t)$ constructed using MST and adding edges with $C^{\tau}_{ij}(t) \geq 0.65$. 
In SI Figures \labelcref{fig:03_correlograms_075_SI,fig:03_correlograms_085_SI}, we show similar correlograms for $S^{\tau}(t)$ constructed using MST and adding edges with $C^{\tau}_{ij}(t) \geq 0.75$ and $C^{\tau}_{ij}(t) \geq 0.85$, respectively. 
Among the two edge-centric discrete Ricci curvatures, FRE shows a higher correlation with three traditional market indicators ($\mu$, volatility, and $\sigma_p$) for both NSE and BSE markets. 
Among the standard network measures, both NE and CE have high correlations with traditional market indicators. 
However, amongst the persistent homology based topological measures, PE outperforms both $L^1$-norm and $L^2$-norms of persistence landscape in terms of correlations with traditional market indicators. 
Hence, we find PE to be a better measure of changes in the topological patterns that appear in the time series of log-returns of daily closing prices in the stock market data. 
Moreover, the persistence landscape and its $L^p$-norms require an appropriate choice of homology group $H_p$ and parameter $k$. 
On the other hand, the PE is a parameter-free measure, which computes only the Shannon entropy of the persistence barcodes. 
Therefore, the persistent entropy could be used as a simple yet powerful measure to better capture the fragility and the dynamical behaviour of a financial system.

Traditionally, econophysicists have employed one of the following filtration approaches to study the correlation structure of a financial system: Minimum Spanning Tree (MST) \cite{bonanno2000taxonomy,bonanno2003topology,nobi2014correlation}, Planar Maximally Filtered Graph (PMFG) \cite{eryiugit2009network,wang2017correlation} or simple threshold networks \cite{kumar2012correlation,nobi2014correlation,chen2020correlation}. 
In all of these methods, the main goal is to extract a backbone comprising of pairs of financial entities with highly relevant correlations among all possible pairs of correlations. 
In an earlier contribution \cite{samal2021frontiers}, some of us have shown that constructing correlation-based threshold networks by starting from MST is a superior framework compared to other traditional network-based approaches to monitor the state of a financial system. 
However, a drawback of this framework, highlighted by Figure \labelcref{fig:03_correlograms} and SI Figures \labelcref{fig:03_correlograms_075_SI,fig:03_correlograms_085_SI}, is that the correlations among the standard network measures and any traditional market indicator ($\mu$, volatility, or $\sigma_p$) can be dependent on the choice of threshold in the network construction. 
In contrast, application of persistent homology for the analysis of financial systems does not require any such choice of threshold, and there is no requirement of extraction of a backbone. 
We compute the persistent homology over the distance matrices $D^{\tau}(t)$ extracted from the corresponding correlation matrices $C^{\tau}(t)$. 
Hence, it is likely to be a better approach for detecting market instabilities in the financial systems compared to the network-based approaches.


\section{Conclusion} \label{sec:conclusion}

In this work, we employed cross-correlation based network analysis and topological data analysis to investigate two Indian stock markets, specifically, 45 stocks from National Stock Exchange (NSE) over a period of 18 years and 77 stocks from Bombay Stock Exchange (BSE) over a period of 23 years. 
We monitored the time series of daily log-returns of stocks from both the NSE and BSE markets by employing several standard network measures, edge-centric discrete Ricci curvatures and persistent homology based topological measures. 
We constructed the correlation-based threshold networks of stocks using the minimum spanning tree (MST) of cross-correlation matrices $C^{\tau}(t)$ and adding edges with correlations $C^{\tau}_{ij}(t) \geq 0.65\ 0.75$ and $0.85$ over the epoch size of $\tau = 22$ days and an overlapping shift of $\Delta{\tau} = 5$ days and ending on trading days $t$. 
By doing this, we ensure that the resulting network $S^{\tau}(t)$ has only a single connected component and it contains only the most relevant edges. 
To the best of our knowledge, this is the first study to employ edge-centric discrete Ricci curvatures for the analysis of the Indian stock market networks to reveal higher-order interactions in the Indian financial system. 

For the topological data analysis of the Indian stock markets, we used the distance matrices $D^{\tau}(t)$ constructed from the cross-correlation matrices $C^{\tau}(t)$ over the epoch size of $\tau = 22$ days and an overlapping shift of $\Delta{\tau} = 5$ days and ending on trading days $t$. 
We built Vietoris-Rips complex to examine the topological features of $D^{\tau}(t)$. 
We used the $L^1$-norm and $L^2$-norm of persistence landscape along with the persistent entropy of persistence barcodes to capture the time evolution of the cross-correlation structures of the Indian stocks. 
Our results show that persistent entropy has the potential to be a simple yet effective measure for capturing the dynamic temporal patterns in Indian stock market data. 

Notably, our work provides a comparison of the network-based analysis with topological data analysis (in particular, persistent homology) for characterizing the correlation structures and identifying the fragility in the two Indian stock markets. 
We showed that, the correlations between a set of standard network measures and the traditional market indicators depend upon the choice of an appropriate threshold used in the construction of correlation-based threshold networks of stocks. 
In contrast, the persistent homology does not consider any particular value of the threshold, but rather considers a range of threshold values to better capture the change in correlation structure of stock markets. 
This underscores the potential utility of persistent homology as a more robust and effective approach to detect instabilities in stock markets (in comparison to the traditional market indicators and network-based approaches).

In conclusion, the methods used in this work can be employed to provide a deeper understanding of the increasingly complex and interconnected financial systems and can be utilized as early warning signals for imminent market crises.


\section*{Acknowledgements}
Areejit Samal would like to acknowledge support from the Max Planck Society, Germany, through the award of a Max Planck Partner Group in Mathematical Biology, and the Department of Atomic Energy (DAE), Government of India, through the Apex project to The Institute of Mathematical Sciences (IMSc), Chennai. Sunil Kumar would like to acknowledge the Science and Engineering Research Board (SERB), Department of Science and Technology, Government of India for providing financial support via MATRICS grant number MTR/2020/000651.


\section*{Competing interests}
The authors declare no competing financial interests.


\appendix
\renewcommand{\theequation}{\thesection.\arabic{equation}}
\section{Traditional market indicators} \label{appendix:market_indicators}

\subsection*{GARCH \textbf{\textit{(p,q)}} process}

Bollerslev \cite{bollerslev1986generalized} introduced the GARCH$(p, q)$ process, which is a generalized ARCH process. In this process, the variable $x_t$, representing a strong white noise process, can be expressed in relation to a time-dependent standard deviation $\sigma_t$. Specifically, $x_t$ can be defined as $\eta_t \sigma_t$, where $\eta_t$ is a random Gaussian process with a mean of $0$ and a variance of $1$. The simplest GARCH process is the GARCH$(1,1)$ process, with Gaussian conditional probability distribution function, given by:
\begin{equation}
\label{eq:garch11}
    \sigma^2_t = \alpha_0 + \alpha_1 x^2_{t-1} + \beta_1 \sigma^2_{t-1}
\end{equation}
where $\alpha_0 > 0$ and $\alpha_1 \geq 0$. $\beta_1$ is an additional control parameter. We can write Eq.(\ref{eq:garch11}) in the form a random multiplicative process as follows:
\begin{equation}
    \sigma^2_t = \alpha_0 + \left( \alpha_1 \eta^2_{t-1} + \beta_1 \right) \sigma^2_{t-1}
\end{equation}
To perform this calculation, we have utilized a built-in function from MATLAB garch (\url{https://in.mathworks.com/help/econ/garch.html}).

\subsection*{Markowitz portfolio optimization}

We employed the Markowitz framework to determine the portfolio with the least amount of risk, considering the risk aversion level of each investor, while maximizing expected returns and minimizing variance. Within this framework, the variance of a portfolio indicates the significance of effectively diversifying investments to minimize overall risk. The Markowitz model minimizes $w'Bw - \phi R' w$, with respect to the normalized weight vector $w$. The covariance matrix $B$ is calculated from the log-returns of stocks, $\phi$ represents the investor's risk appetite, and $R'$ denotes the expected return of the assets. We imposed a short-selling constraint, i.e., set $\phi = 0$ as well as $w_i \geq 0$, in order to find a convex combination of stock returns that yields the minimum risk portfolio. To perform this calculation, we have utilized a built-in function in MATLAB Portfolio (\url{https://in.mathworks.com/help/finance/portfolio.html}).

\section{Standard network measures} \label{appendix:network_measures}

In this study, every correlation-based threshold network of stocks is considered as an undirected and weighted graph, denoted by $G(V, E)$, where $V$ represents the set of nodes, $E$ denotes the set of edges, and each edge $(i, j) \in E$ has a weight $w_{ij}$. Since, these graphs are also weighted graphs, the edge-weights can either be the strength or the distance between two nodes. The appropriate choice of weight to be used depends upon the network measure used to analyze the weighted graph and the corresponding computation of it. We provide a list of standard network measures and discrete Ricci curvatures used in our analysis in SI Table \ref{tab:network_measures}, categorized by its type and choice of the weight to be used. Using cross-correlations among the stocks, we constructed undirected and weighted networks. In these networks, we employ the absolute correlation value, $\scriptstyle |C_{ij}^{\tau}(t)|$, between two stocks $i$ and $j$, to represent the edge strength, which would vary from 0 to 1. And, we use the ultrametric distance $\scriptstyle D_{ij}^{\tau} (t) = \sqrt{2(1-C_{ij}^{\tau}(t))}$ to represent the distance of the edge between stocks $i$ and $j$, which can take value between 0 and 2.
In this study, we have analyzed the structure of the weighted and undirected correlation-based threshold networks using the following measures:

\begin{itemize}
    \item \emph{Edge count:} The cardinality of set $E$, denoted as $m = |E|$, specifies the edge count of the network.

    \item \emph{Edge density:} The ratio of number of edges $m$ and the number of all possible edges gives the edge density of a network. In an undirected case, with no self-edges, the maximum number of edges is $\frac{n(n-1)}{2}$. Thus, the edge density can be calculated using the expression $\frac{2m}{n(n-1)}$.
   
    \item \emph{Average degree:} A degree of a particular node in the network is given by the number of nodes connected to it. The average of all the degree values in the network is called the average degree of the network. It is given as $\langle k \rangle = \frac{2m}{n}$, where $m$ is the number of edges, and $n$ is the number of nodes given by the cardinality of set $V$, denoted as $n = |V|$.
    
    \item \emph{Clique number:} The largest number of nodes that form a fully connected subgraph within a larger network is called as the clique number.
    
    \item \emph{Network entropy:} This quantity measures the extent of disorder or complexity in the connectivity patterns of a network \cite{sole2004information}. To calculate the network entropy, the diversity of edge distribution is measured using the remaining degree distribution $q_k$, which represents the probability of a node having excess degree $k$. This remaining degree distribution $q_k$ is given by $q_k = \frac{(k+1) p_{k+1}}{\langle k \rangle}$, where $p_{k+1}$ represents the probability of a node having degree $k+1$ and $\langle k \rangle$ is the average degree of the network. The network entropy $H(q)$ is then calculated as:
    \begin{equation}
        H(q) = - \sum_k q_k log(q_k).
    \end{equation}
    
    \item \emph{Average weighted degree:} The weighted degree of a particular node is given by the sum of the weights on the edges connected to it in the network \cite{barrat2004architecture}. It is given by $k_i^w = \sum_j w_{ij}$, where $j$ is a neighboring node of $i$, and $w_{ij}$ is the weight of the edge between $i$ and $j$. The average of this quantity over all nodes in the network is called the average weighted degree, which is given by:
    \begin{equation}
        \langle k_w \rangle = \sum_i k_i^w = \frac{2m_w}{n}
    \end{equation}
    where $m_w = \sum_{(i,j)} w_{ij}$ is the sum of all edge weights in the network.
    
    \item \emph{Global assortativity:} A preference of a node in the network to attach to other nodes with similar characteristics is referred to as assortative mixing \cite{newman2002assortative, newman2003mixing}. If a significant fraction of edges in the network are between nodes of the same type, then that network is said to be assortative. In contrast, if a significant fraction of edges in the network are between nodes of the different type, then that network is disassortative. Quite often, assortativity is computed in terms of nodes' degrees, where it measures the degree correlations between the nodes of an unweighted network. In case of degree-based assortative networks, the nodes of comparable degree tend to connect with each other, i.e., low degree nodes tend to connect with other low degree nodes, and high degree nodes tend to connect to other high degree nodes. In degree-based disassortative networks, high degree nodes tend to connect to low degree nodes and low degree nodes to high degree nodes. Notably, the definition of degree assortativity has also been extended to the weighted networks \cite{leung2007weighted}. The global assortativity of a weighted network is given by:
    \begin{equation}
        r_w = \frac{\frac{1}{m_w} \sum_{(i,j)} w_{ij} k_i k_j - \left[ \frac{1}{2m_w} \sum_{(i,j)} w_{ij} (k_i + k_j) \right]^2}{\frac{1}{2m_w} \sum_{(i,j)} w_{ij} (k_i^2 + k_j^2) - \left[ \frac{1}{2m_w} \sum_{(i,j)} w_{ij} (k_i + k_j) \right]^2}
    \end{equation}
    where $m_w$ is the sum of weights assigned to all the edges in the network, $k_i$ and $k_j$ are the degrees of node $i$ and $j$, $w_{ij}$ is the weight of the edge between nodes $i$ and $j$, and all the summation runs over all the edges $(i,j)$ in the weighted network. The quantity $r_w$ varies between $-1 \leq r_w \leq 1$, i.e. for $r_w > 0$ the network is assortative by degree, for $r_w = 0$ the network is neutral, and for $r_w < 0$ the network is disassortative by degree.
    
    \item \emph{Average clustering coefficient:} The clustering coefficient is a measure of the local density of connections in the immediate neighborhood of a particular node in the network. In other words, it quantifies the degree to which the neighbors of a node are also connected to each other. This quantity can be generalized for computing the clustering coefficient of a node in a weighted network, which takes into account the edge weights for edges between the neighbors of that node \cite{onnela2005intensity}. The clustering coefficient of a node $i$ in the weighted network is given by:
    \begin{equation}
        C_i = \frac{2}{k_i(k_i - 1)} \sum_{j,k} (w_{ij} w_{ik} w_{jk})^{1/3}
    \end{equation}
    where $j$ and $k$ are the neighbors of node $i$, $k_i$ is the degree of node $i$, $w$ represents edge weight between nodes in the subscript, and the summation runs over all such pairs of neighbors of node $i$. Its value ranges from 0 to 1, with higher values indicating a greater tendency for its neighbors to form a densely connected cluster. Hence, the average clustering coefficient of a network can be calculated by taking the average of the clustering coefficients for all its nodes.
    
    \item \emph{Modularity:} It is a measure of division or community structure in the network. It quantifies the extent to which nodes in the network tend to be grouped together into densely connected clusters or communities, with relatively fewer connections between the pairs of nodes from different clusters. The modularity for a weighted network \cite{girvan2002community, blondel2008fast} can be defined as:
    \begin{equation}
    Q = \frac{1}{2m_w} \sum_{i \neq j \in V } \left[ w_{ij} - \frac{m_i^w m_j^w}{2m_w} \right] \delta (c_i, c_j)
    \end{equation}
    where $m_i^w$ and $m_j^w$ represent the sum of weights of edges attached to nodes $i$ and $j$, respectively, $c_i$ and $c_j$ are the communities of nodes $i$ and $j$, respectively, and $m_w$ is the sum of all edge weights in the network.
    
    \item \emph{Diameter:} A walk in a network is any route that runs from node to node along the edges. A walk can visit a node or can run along an edge or can intersect itself multiple times. In contrast, a walk that does not intersect itself is called a path. The shortest walk that traverses smallest number of edges between given two nodes is called a shortest path between them. The maximum value of all existing shortest paths between all the pairs of nodes in the network, i.e., $\max \left\lbrace d(i,j) \forall i,j \in V \right\rbrace$, gives the diameter of the network.
    
    \item \emph{Global reaching centrality (GRC):} It is a measure of the ability of a node to reach other nodes in the network through shortest paths. It was proposed as a way to identify important nodes in a network based on their global influence and the measure can also provide information on hierarchical organization in networks \cite{mones2012hierarchy}. It was introduced only for unweighted and undirected networks, but it can be extended to weighted networks as follows:
    \begin{equation}
        GRC = \frac{\sum_{i \in V} \left[ C^{max} - C(i) \right]}{n-1}
    \end{equation}
    where $C(i)$ is the local reaching centrality (LNC) of node $i$, and $C^{max}$ is the maximum value of LNC for all nodes in the network. The LNC for node $i$ is given by:
    \begin{equation}
        C(i) = \frac{1}{n-1} \sum_{j \in V \symbol{92} \lbrace i \rbrace} \frac{1}{d(i,j)} 
    \end{equation}
    where $d(i,j)$ is the shortest path length between nodes $i$ and $j$ in the network.
    
    \item \emph{Communication efficiency:} It is a measure of the efficiency of communication between pairs of nodes in the network \cite{latora2001efficient}. It quantifies the ease with which information can flow between the nodes of the network, taking into account both the network structure and the shortest path lengths between the nodes. The communication efficiency of a network is defined as:
    \begin{equation}
        \mu_c = \frac{1}{n(n-1)} \sum_{i \neq j \in V} \frac{1}{d(i,j)}
    \end{equation}
    where $n$ is the number of nodes and $d(i,j)$ is the shortest path length between nodes $i$ and $j$ in the network.
\end{itemize}
\section{Discrete Ricci curvatures} \label{appendix:discrete_ricci_curvatures}

Ricci curvature in a fundamental concept in Riemannian geometry which is defined for smooth manifolds \cite{jost2008riemannian}. Subsequently, the concept has been extended to discrete structures including networks. While developing meaningful notions of Ricci curvature for discrete objects, it is necessary to recall that classical Ricci curvature in smooth manifolds captures two geometric properties namely, volume growth and dispersion of geodesics. Importantly, a discretization of Ricci curvature may only be able to capture some of the key properties of classical notion. Therefore, different discrete notions of Ricci curvatures (proposed in the literature), capture different aspects of the classical notion, and thus, may reveal different structural properties of complex networks \cite{samal2018comparative}.

In smooth manifolds, the classical definition of Ricci curvature is linked to vectors, whereas in the context of graphs or networks, Ricci curvature is inherently linked to the edges of the network. Therefore, discrete graph Ricci curvatures are edge-based measures which can be used to identify influential edges in the network, in contrast to commonly used node-based centrality measures such as degree, closeness, betweenness and eigenvector centrality \cite{samal2018comparative}. Next, we describe two notions of discrete Ricci curvature that were utilized for analyzing stock market networks here.


\subsection*{Forman-Ricci curvature}
Forman proposed a discretization of Ricci curvature which captures the geodesic dispersal property of the classical Ricci curvature \cite{forman2003bochner}. It was originally introduced in the general framework of weighted $CW$ cell complexes, and was later adapted for the analysis of complex networks \cite{samal2018comparative, sreejith2016forman, sreejith2017systematic}. In the network context, the Forman-Ricci curvature of an edge measures the information spread at the ends of that edge. Larger the spread of information at the ends of an edge, the more negative value of the Forman-Ricci curvature for that edge. The extension of Forman's discretization to undirected and weighted networks, can accommodate the weights assigned to both edges and nodes in the network \cite{sreejith2016forman}. Specifically, in an undirected and weighted network, Forman-Ricci curvature of an edge $e$ between nodes $i$ and $j$ can be defined as: 
\begin{equation}
    \mathbf{F}(e) = w_{e} \left( \frac{w_i}{w_{e}} + \frac{w_j}{w_{e}} - \sum_{e_i \sim e, e_j \sim e} \left[ \frac{w_i}{\sqrt{w_e w_{e_i}}} + \frac{w_j}{\sqrt{w_e w_{e_j}}} \right] \right)
\end{equation}
where $w_e$ denotes the weight of the edge $e$ between nodes $i$ and $j$, $w_i$ and $w_j$ are weights of the nodes $i$ and $j$, respectively. Moreover, $e_i \sim e$ and $e_j \sim e$ denote the set of edges incident on nodes $i$ and $j$, respectively, excluding the edge $e$ between nodes $i$ and $j$. Finally, the $w_{e_j}$ and $w_{e_i}$ represent the weights of the edges $e_i$ and $e_j$, respectively. In this work, while computing Forman-Ricci curvature of edges in the correlation-based stock market networks, the edge weight is taken to be the distance measure given by Eq.(\ref{eq:distance_measure}), and the node weight is taken to be 1. Further, we computed the average Forman-Ricci curvature of all edges in the stock market networks.


\subsection*{Ollivier-Ricci curvature}

A discretization of classical Ricci curvature, which captures the geometric property of volume growth, was proposed by Ollivier \cite{ollivier2007ricci, ollivier2009ricci}. This notion of discrete Ricci curvature, namely Ollivier-Ricci curvature, has been widely used to analyze complex networks \cite{samal2018comparative, lin2011ricci, ni2019community, sandhu2016ricci, samal2021frontiers, samal2021rsos}. In an undirected network $G$, the Ollivier-Ricci curvature of an edge $e$ between nodes $i$ and $j$ is given by:
\begin{equation}
    \mathbf{O}(e) = 1 - \frac{W_1 (m_i, m_j)}{d(i,j)}
\end{equation}
where $d(i,j)$ is the shortest path length between nodes $i$ and $j$, both in weighted or unweighted case, $m_i$ and $m_j$ are discrete probability measures defined on nodes $i$ and $j$, respectively, and Wasserstein distance $W_1(m_i, m_j)$ \cite{vaserstein1969markov} denotes the transportation distance between $m_i$ and $m_j$ and is given by:
\begin{equation}
    W_1 (m_i, m_j) = \inf_{\mu_{i,j} \in \prod (m_i, m_j)} \sum_{(i',j') \in V \times V} d(i',j') \mu_{i,j} (i',j')
\end{equation}
where $\prod (m_i, m_j)$ is the set of probability measures $\mu_(i,j)$ that satisfy:
\begin{equation}\label{eq:possible_trans}
    \sum_{j' \in V} \mu_{i,j} (i',j') = m_i (i'), \sum_{i' \in V} \mu_{i,j} (i',j') = m_j (j')
\end{equation}
where $V$ is the set of nodes in the network $G$. Eq.(\ref{eq:possible_trans}) gives all possible transportations of measures $m_i$ and $m_j$, and the $W_1(m_i,m_j)$ is the minimal cost of transporting $m_i$ to $m_j$. In computations of Ollivier-Ricci curvature, it is necessary to specify the probability distribution $m_i$ beforehand. In this work, it is taken to be uniform over the neighboring nodes of node $i$ \cite{lin2011ricci}.
While computing Ollivier-Ricci curvature of edges in the network, the weight of each edge is chosen to be the distance measure given by Eq.(\ref{eq:distance_measure}). In this work, we computed the average Ollivier-Ricci curvature of all edges in the stock market networks.


\section{Persistent homology} \label{appendix:persistent_homology}

In this section, we will briefly introduce and discuss the concepts essential for the study of persistent homology and corresponding TDA-inspired measures used in this study.


\subsection*{Simplicial complex}
A simplicial complex $K$ is a collection of finite number of simplices, where $p$-dimensional simplex ($p$-simplex) $\alpha$ in $K$ is the convex hull of its $p+1$ affinely independent points \cite{zomorodian2004computing}. In this context, a single point corresponds to 0-simplex, an edge connecting two points corresponds to 1-simplex, a triangle made by 3 edges corresponds to 2-simplex, and a tetrahedron formed by 4 triangles corresponds to 3-simplex. Given a $p$-simplex $\alpha$ in $K$, the face $\gamma$ of the simplex $\alpha$ is determined by a subset of the points in $\alpha$ with cardinality $\leq$ $p+1$. Hence, the dimension of simplicial complex $K$ is the maximum dimension of its constituent simplices. In a formal sense, a simplicial complex $K$ is a set of simplices that satisfies the following conditions: 
\begin{enumerate}
\item Any face $\gamma$ of a simplex $\alpha$ in $K$ is also an element of $K$.
\item If intersection of two simplices $\alpha$ and $\beta$ in $K$ is non-empty and defined by $\gamma$, then $\gamma$ is a common face of $\alpha$ and $\beta$.
\end{enumerate}
For a deeper understanding of simplicial complexes, we refer the interested reader to a standard texts in algebraic topology (see e.g., \cite{munkres2018elements}).


\subsection*{Persistent homology of a simplicial complex}

As defined in the previous section, for a given a simplicial complex $K$, the subcomplex of $K$ is a simplicial complex $K_i \subset K$. A filtration of a simplicial complex $K$ is a nested sequence of its sub-complexes
\begin{equation}
    \varnothing = K_0 \subseteq K_1 \subseteq \ldots \subseteq K_n = K
\end{equation}
Persistent homology groups can be defined for a simplicial complex $K$ with a given filtration as follows. A base field $F$ is fixed, and the collection of all oriented $p$-simplices in $K$ generates a free group $C_p$ over $F$, which is called the $p^{\text{th}}$-chain group. A $p$-chain is an element of $C_p$ and is represented by a finite formal sum.
\begin{equation}
    C_p = \sum^N_{i=1} c_i \alpha_i
\end{equation}
where the coefficients $c_i$ represent elements from the base field $F$, while $\alpha_i$ are the oriented $p$-simplices belonging to the simplicial complex $K$. The group structure of $C_p$ is provided by component-wise addition, with the identity element being the sole $p$-chain having all coefficients $c_i$ equal to zero. The inverse of a $p$-simplex $\alpha$ in $C_p$ is represented as $-\alpha$ when $\alpha$ is given an opposite orientation. In order to define the persistent homology groups, a mapping called the boundary operator $\partial_p$ is utilized, which maps from $C_p$ to $C_{p-1}$.

For an oriented $p$-simplex $\alpha = [x_0, x_1, \ldots, x_p]$, we define $\partial_p$ as follows:
\begin{equation}
    \partial_p (\alpha) = \sum^p_{i=0} (-1)^i [x_0, \ldots, \hat{x}_i, \ldots, x_p]
\end{equation}
where $[x_0, \ldots, \hat{x}_i, \ldots, x_p]$ denotes a $(p-1)$-face of $\alpha$ obtained by removing the vertex $x_i$. The right-hand side of the above equation belongs to $C_{p-1}$, since it is a linear combination of $(p-1)$-simplices. By linearity, the definition of $\partial_p$ can be extended to all elements of $C_p$. Therefore, the boundary operators have the fundamental property:
\begin{equation}
    \partial_p \circ \partial_{p+1} = 0
\end{equation}
i.e., the boundary of boundary does not exist. A $p$-chain is called as a cycle if its boundary is zero. This results into the group of $p$-cycles, denoted by $Z_p$, being defined as the kernel of the boundary operator $\partial_p$. It consists of the set of elements in $C_p$ that are mapped to 0 in $C_{p-1}$ by $\partial_p$:
\begin{equation}
    Z_p = \text{Ker}(\partial_p) = \lbrace c \in C_p \mid \partial_p(c) = 0 \rbrace.
\end{equation}
A $p$-cycle that lies in the image of the boundary operator $\partial_{p+1}$ is called a $p$-boundary. The group of $p$-boundaries is denoted by $B_p$ and is a subgroup of $Z_p$:
\begin{equation}
    B_p = \text{Image}(\partial_{p+1}) = \lbrace c \in C_p \mid \exists~b \in C_{p+1} , \partial_{p+1}(b) = c \rbrace.
\end{equation}
Thus, the $p$-homology group is defined as:
\begin{equation}
    H_p(K) = \frac{Z_p(K)}{B_p(K)}.
\end{equation}
It should be noted that $H_p$ forms a vector space over the field $F$, and the $p$-Betti number $\beta_p$ is given by the dimension of the homology group $H_p$. In simpler terms, $\beta_p$ provides an informal indication of the number of holes present in the $p$-homology group. Notably, $\beta_0$ gives the number of connected components, $\beta_1$ gives the number of loops or holes, and $\beta_2$ gives the number of voids or cavities.

In the filtration of the simplicial complex $K$, each subcomplex $K_i \subseteq K$ is associated with an index $i$. Furthermore, there exists a corresponding $p$-chain, $p$-boundary operator, $p$-boundaries, and $p$-cycles for each subcomplex $K_i$. Let us use $Z^i_p$ to denote the $p$-cycles of $K_i$, and $B^i_p$ to denote the $p$-boundaries of $K_i$. With this notation, we can define the $j$-persistent $p$-homology of $K_i$ as follows:
\begin{equation}
    H^{i,j}_p = \frac{Z^i_p}{( B^{i+j}_p \cap Z^i_p)}
\end{equation}
and the corresponding $j$-persistent $p$-Betti number as:
\begin{equation}
    \beta^{i,j}_p = dim(H^{i,j}_p).
\end{equation}
If a $p$-homology class is not in the image of the map induced on $p$-homology by the inclusion $K_{i-1} \subseteq K_i$, then we say that it is born at $K_i$. Additionally, if $p$-homology class is born at $K_i$, then it dies entering $K_{i+j}$ if it becomes the boundary of a $(p + 1)$-chain in $K_{i+j}$. The persistent homology group $H^{i,j}_p$ contains information about the $p$-homology classes that originate at the filtration index $i$ and persist until the index $i+j$. Persistent homology provides a means to characterize the birth and death of each $p$-hole across the filtration and to quantify the persistence of such holes. This enables the identification of important topological features throughout the filtration process. For more details on persistent homology, see e.g. \cite{carlsson2009topology, otter2017roadmap, pun2018persistent}.

\subsection*{Vietoris-Rips complex}

Consider a set of points $S$ in a finite metric space with some distance function. Such spaces are also called point clouds in the TDA literature. We are interested in understanding the topological features of this space, such as existence of number of connected components, loops, and voids. Persistent homology can be used to capture these features using advanced algebraic structures. However, it is difficult to compute homology of an arbitrary topological space $X$. Hence, the simplicial complexes whose homology can appropriately approximate the homology of that space are utilized. The \v{C}ech complex is one such classical example of a simplicial complex from algebraic topology \cite{edelsbrunner2022computational}. From a practical perspective, the \v{C}ech complex is computationally very expensive. To address this issue, an approximation of the \v{C}ech complex, called the Vietoris-Rips complex or simply the Rips complex \cite{ghrist2008barcodes, kerber2013approximate, dantchev2012efficient} can be considered.

For a non-negative real number $\epsilon$, the Rips complex $R_{\epsilon}(S)$ with radius $\epsilon$ is a simplicial complex whose vertices are the points of $S$, and whose simplices are all finite subsets of $S$ with pairwise distances $\leq$ $\epsilon$. Formally, Rips complex $R_{\epsilon}(S)$ with radius $\epsilon$ is defined as:
\begin{equation}
\label{eq:rips_complex}
    R_{\epsilon}(S) = \left \lbrace \sigma \subseteq S~\vert~d(p,q) \leq 2\epsilon~\forall~p,q \in \sigma \right\rbrace 
\end{equation}
where $d(p,q)$ is the distance between points $p$ and $q$. Hence, in Rips complex, the filtration starts with an empty simplicial complex consisting of only the set of points $S$, and gradually edges, triangles, tetrahedra, and so on, are added as the parameter $\epsilon$ increases. If two points are within a distance of $\epsilon$, then they are connected by an edge; if three points are all within a distance of $\epsilon$, then they form a filled-in triangle; and if four points are all within a distance of $\epsilon$, then they form a filled-in tetrahedron, and so on.

Moreover, determining whether $\sigma \subseteq S$ belongs to $R_{\epsilon}(S)$ in Eq.(\ref{eq:rips_complex}) is the same as determining if the maximum pairwise distance between any pair of points $p$ and $q$ in $\sigma$ is less than or equal to $2\epsilon$. This fact makes it very easy to compute the Rips complex, because to construct the complex one has only to check for pairwise distances. For a detailed discussion on Rips complex, see e.g.  \citep{vietoris1927hoheren, edelsbrunner2022computational}.

\subsection*{Persistence barcode and persistent entropy}

As stated above, we use persistent homology to measure several topological features of the space $S$, such as number of components, loops, and voids. The lifetime of such features can be visualized using a finite collection of intervals called the persistence barcode. The $p^{\text{th}}$-barcode diagram for a given filtration of a finite simplicial complex $K$ gives a graphical summary of the birth and death of $p$-holes across the filtration \cite{ghrist2008barcodes}. A horizontal line in the $p^{\text{th}}$-barcode diagram of $K$ is referred to as a barcode. A barcode that begins at a $x$-axis value of $\epsilon_i$ and ends at a $x$-axis value of $\epsilon_j$ represents a $p$-hole in $K$ whose birth and death filtration values are $\epsilon_i$ and $\epsilon_j$, respectively. The number of barcodes between $\epsilon_i$ and $\epsilon_j$ in the diagram is precisely the $p$-Betti number $\beta^{i,j-i}$, i.e., the dimension of the persistent homology group $H^{i,j-i}_p$.

While the persistence barcodes are useful for visualization of the persistence of toplogical features over various resolutions, they aren’t suitable for statistical analysis. To quantify the sequence of birth-death information of the topological features, we evaluate a topological measure called the persistent entropy \cite{chintakunta2015entropy, rucco2016characterisation, atienza2020stability}. It is the Shannon entropy of the persistence barcode. Consider a probability distribution $\lbrace p_i \rbrace$ with $i = 1, 2, \dots, n$ and $0 \leq p_i \leq 1$, which satisfy $\sum_{i=1}^n p_i = 1$. The Shannon entropy of this distribution is given by:
\begin{equation}
    H = \sum_{i=1}^n -p_i log(p_i)
\end{equation}
where $0 \leq H \leq log(n)$. The quantity $H$ reaches the maximum value of $log(n)$ only if $p_i = 1/n$. A similar definition of entropy can be extended to persistence barcode, where it measures the entropy of the bars. Consider a barcode $\lbrace (b_i, d_i) \rbrace$, where $b_i$ and $d_i$ are birth and death filtration values of $i^{\text{th}}$ bar. We compute the length of each bar by using a simple expression, $l_i = d_i - b_i$. Now, we can express the barcode as a probability distribution $e_i = l_i / L$, where total length $L = \sum_{i=1}^n l_i$ is a summation over all the bars. The persistent entropy of this distribution can be given by:
\begin{equation}
    E = \sum_{i=1}^n -e_i log(e_i)
\end{equation}
The value of $E$ will be greater if there are many bars in the barcode with balanced length.

\subsection*{Persistence diagram and persistence landscape}

A persistence diagram is the set of all $p$-dimensional birth-death pairs forming a multiset of points in $\mathbb{R}^2$, and it gives the same information as a barcode \cite{edelsbrunner2022computational}.  In this diagram, birth filtration value $b_i$ is plotted on $x$-axis and corresponding death filtration value $d_i$ is plotted on $y$-axis, since $d_i$ of a $i^{\text{th}}$ $p$-hole succeeds $b_i$, and points further away from the $x = y $ line denote relatively robust $p$-holes. Persistence diagrams are known to be stable under perturbations from noisy data \cite{cohen2005stability}. Additionally, the metric spaces of persistence diagrams can be defined using a distance function, such as Wasserstein distance \cite{mileyko2011probability, turner2014frechet}. However, these metric spaces are incomplete because of their multiset nature, and are not suitable for performing any statistical analysis. Hence, quite often the space of persistence diagrams is mapped onto the spaces which are more suitable for performing statistical analysis and machine learning.

There are many tools to perform a mapping of spaces as mentioned above, but one particular tool is of our interest, that is the persistence landscape \cite{bubenik2015statistical}. It is a sequence of real-valued functions that summarize the crucial information contained in persistence diagrams. A persistence landscape can be obtained from a persistence diagram as follows. A persistence diagram is rotated by $45^{\circ}$ clockwise, so that the diagonal becomes the $x$-axis. Then, considering each point $(b_i, d_i)$ of persistence diagram as a vertex of an isosceles right-angle triangle, the triangles are drawn accordingly. Next, from resulting set of triangles, each individual tent function is derived. For example, the third landscape function corresponds to the point-wise third largest value among all the triangles. Mathematically, for each birth-death pair $\lbrace (b_i, d_i) \rbrace$ in a persistence diagram, there exists a piece-wise linear function $\Lambda_i : \mathbb{R} \rightarrow [0, \infty)$ given by:
\begin{equation}
    \Lambda_i (t) = 
    \begin{cases}
      t - b_i, & t \in \left[b_i, \frac{b_i + d_i}{2} \right] \\
      d_i - t, & t \in \left[\frac{b_i + d_i}{2}, d_i \right] \\
      0 & \text{otherwise.}
    \end{cases}
\end{equation}
The persistence landscape is defined as $\lambda_k (t)$ which is the $k^{\text{th}}$ largest value among the set of $\lbrace \Lambda_i \rbrace$. In our work, we set $k = 1$. Since, the persistence landscape form a subset Banach space, $L^p$-norms of persistence landscape can be computed, which is not possible with persistence diagrams. Here, we compute $L^1$-norm and $L^2$-norm of persistence landscape $\lambda_1(t)$.

\bibliographystyle{unsrt}
\bibliography{references.bib}

\clearpage
\setcounter{table}{0}
\renewcommand{\thetable}{S\arabic{table}}
\setcounter{figure}{0}
\renewcommand{\thefigure}{S\arabic{figure}}%
\section*{Supplementary Information (SI)} \label{appendix:supplementary_info}

\begin{figure}[!ht]
    \centering
    \includegraphics[width=\textwidth,keepaspectratio]{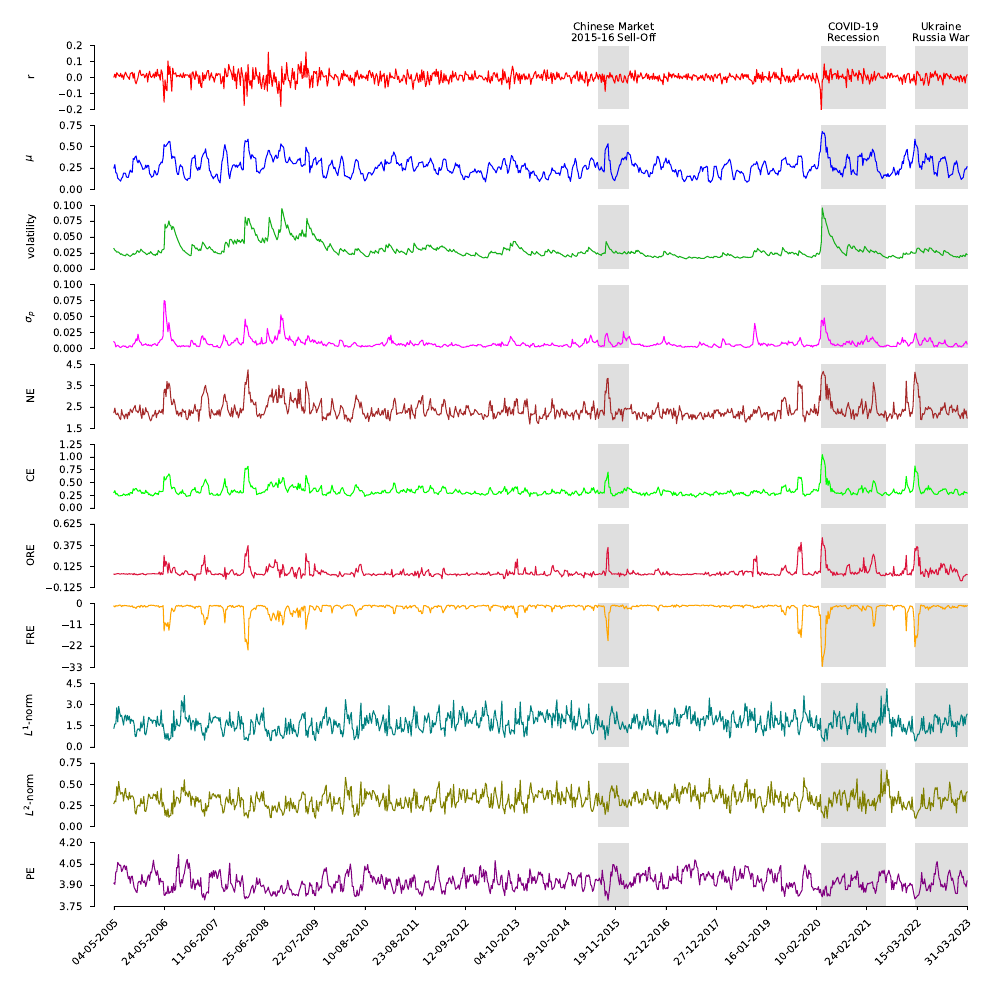}
    \caption{\textbf{Evolution of traditional market indicators, standard network measures, edge-centric discrete Ricci curvatures, and persistent homology based topological measures for the National Stock Exchange (NSE).} We evaluate standard network measures and edge-centric discrete Ricci curvatures on correlation-based threshold networks $S^{\tau}(t)$ of stocks from NSE, constructed by adding edges with correlation $C^{\tau}_{ij}(t) \geq 0.75$ to the MST. We compute topological measures from the persistent homology of the distance matrices $\scriptstyle D^{\tau}(t) = \sqrt{2 (1 - C^{\tau}(t))}$ corresponding to the epochs of size $\tau = 22$ days and an overlapping shift of $\Delta{\tau} = 5$ days. From top to bottom, we compare the index log-returns ($r$), mean market correlation ($\mu$), volatility of the market index $r$ estimated using GARCH(1,1) process, risk ($\sigma_p$) corresponding to the minimum risk Markowitz portfolio of all the stocks in the market, network entropy (NE), communication efficiency (CE), average Ollivier-Ricci edge curvature (ORE), average Forman-Ricci edge curvature (FRE), $L^1$-norm and $L^2$-norm of persistence landscape and persistent entropy (PE). The grey-shaded regions indicate the epochs of important events in the NSE market: Chinese market 2015-16 sell-off, COVID-19 Recession and Ukraine Russia war.}
    \label{fig:01_nse_time_series_075_SI}
\end{figure}
\begin{figure}[ht]
    \centering
    \includegraphics[width=\textwidth,keepaspectratio]{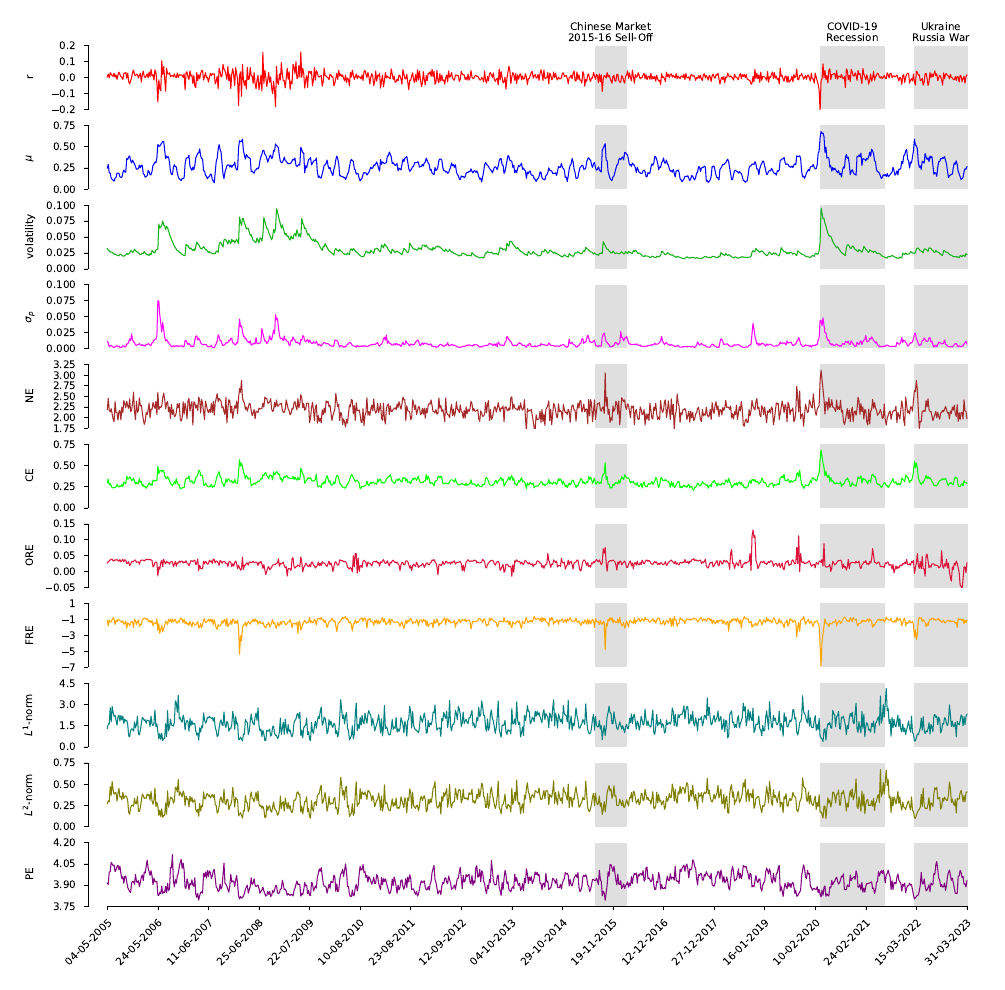}
    \caption{\textbf{Evolution of traditional market indicators, standard network measures, edge-centric discrete Ricci curvatures, and persistent homology based topological measures for the National Stock Exchange (NSE).} We evaluate standard network measures and edge-centric discrete Ricci curvatures on correlation-based threshold networks $S^{\tau}(t)$ of stocks from NSE, constructed by adding edges with correlation $C^{\tau}_{ij}(t) \geq 0.85$ to the MST. We compute topological measures from the persistent homology of the distance matrices $\scriptstyle D^{\tau}(t) = \sqrt{2 (1 - C^{\tau}(t))}$ corresponding to the epochs of size $\tau = 22$ days and an overlapping shift of $\Delta{\tau} = 5$ days. From top to bottom, we compare the index log-returns ($r$), mean market correlation ($\mu$), volatility of the market index $r$ estimated using GARCH(1,1) process, risk ($\sigma_p$) corresponding to the minimum risk Markowitz portfolio of all the stocks in the market, network entropy (NE), communication efficiency (CE), average Ollivier-Ricci edge curvature (ORE), average Forman-Ricci edge curvature (FRE), $L^1$-norm and $L^2$-norm of persistence landscape and persistent entropy (PE). The grey-shaded regions indicate the epochs of important events in the NSE market: Chinese market 2015-16 sell-off, COVID-19 Recession and Ukraine Russia war.}
    \label{fig:01_nse_time_series_085_SI}
\end{figure}
\begin{figure}[!ht]
    \centering
    \includegraphics[width=\textwidth,keepaspectratio]{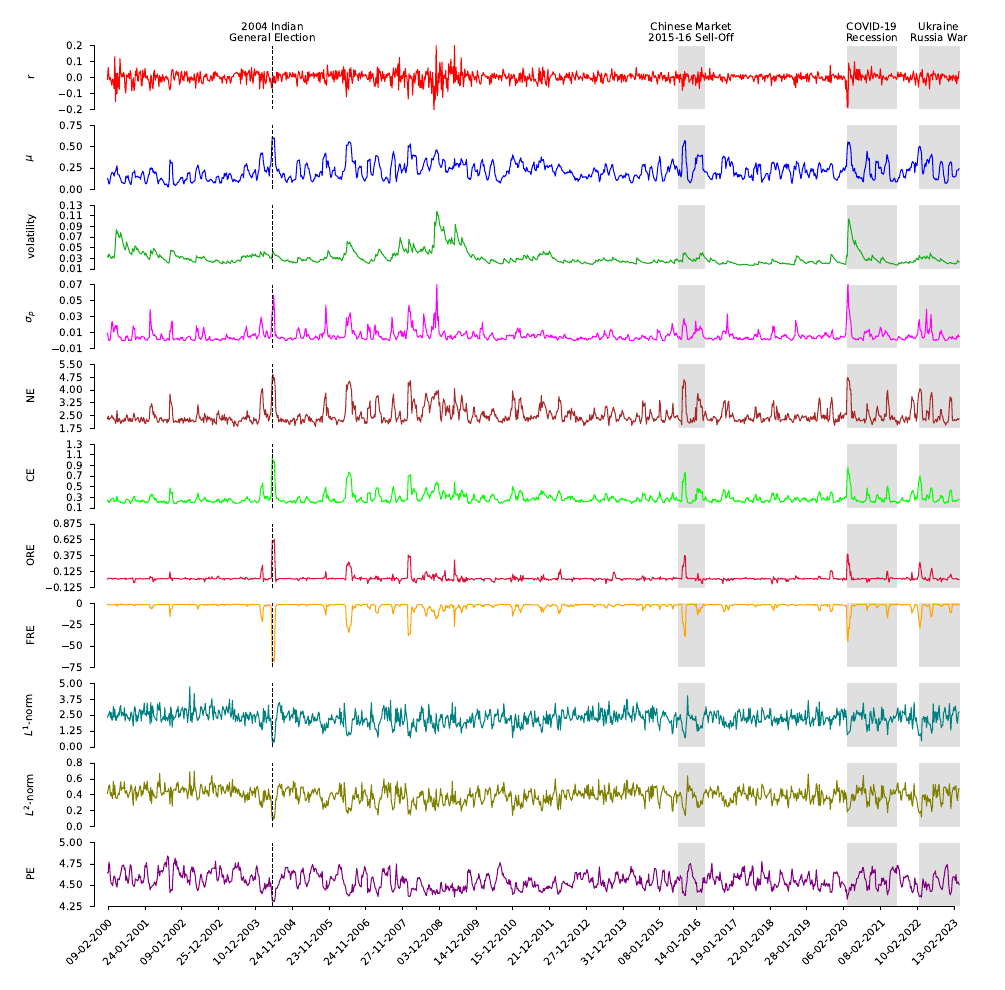}
    \caption{\textbf{Evolution of traditional market indicators, standard network measures, edge-centric discrete Ricci curvatures, and persistent homology based topological measures for the Bombay Stock Exchange (BSE).} We evaluate standard network measures and edge-centric discrete Ricci curvatures on correlation-based threshold networks $S^{\tau}(t)$ of stocks from BSE, constructed by adding edges with correlation $C^{\tau}_{ij}(t) \geq 0.75$ to the MST. We compute topological measures from the persistent homology of the distance matrices $\scriptstyle D^{\tau}(t) = \sqrt{2 (1 - C^{\tau}(t))}$ corresponding to the epochs of size $\tau = 22$ days and an overlapping shift of $\Delta{\tau} = 5$ days. From top to bottom, we compare the index log-returns ($r$), mean market correlation ($\mu$), volatility of the market index $r$ estimated using GARCH(1,1) process, risk ($\sigma_p$) corresponding to the minimum risk Markowitz portfolio of all the stocks in the market, network entropy (NE), communication efficiency (CE), average Ollivier-Ricci edge curvature (ORE), average Forman-Ricci edge curvature (FRE), $L^1$-norm and $L^2$-norm of persistence landscape and persistent entropy (PE). The vertical dashed lines and grey-shaded regions indicate the epochs of important events in the BSE market: 2004 Indian general election, Chinese market 2015-16 sell-off, COVID-19 Recession and Ukraine Russia war.}
    \label{fig:02_bse_time_series_075_SI}
\end{figure}
\begin{figure}[ht]
    \centering
    \includegraphics[width=\textwidth,keepaspectratio]{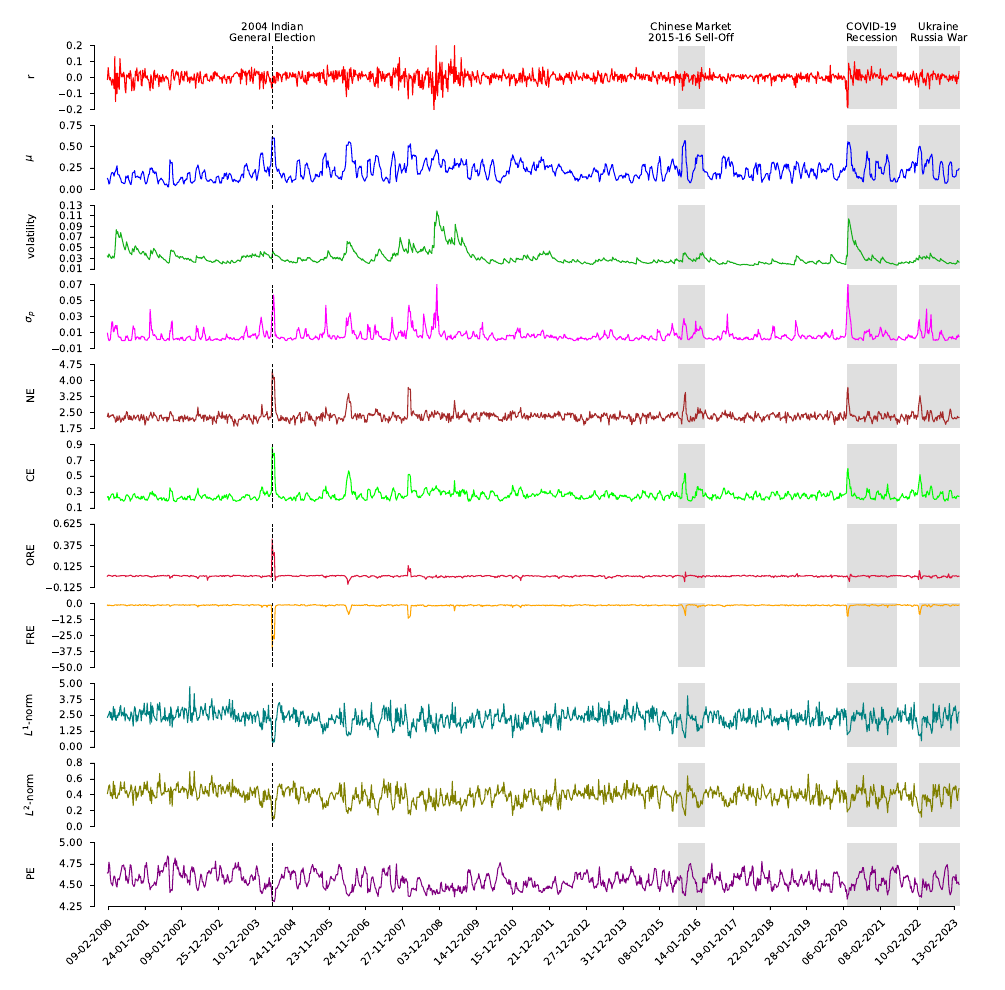}
    \caption{\textbf{Evolution of traditional market indicators, standard network measures, edge-centric discrete Ricci curvatures, and persistent homology based topological measures for the Bombay Stock Exchange (BSE).} We evaluate standard network measures and edge-centric discrete Ricci curvatures on correlation-based threshold networks $S^{\tau}(t)$ of stocks from BSE, constructed by adding edges with correlation $C^{\tau}_{ij}(t) \geq 0.85$ to the MST. We compute topological measures from the persistent homology of the distance matrices $\scriptstyle D^{\tau}(t) = \sqrt{2 (1 - C^{\tau}(t))}$ corresponding to the epochs of size $\tau = 22$ days and an overlapping shift of $\Delta{\tau} = 5$ days. From top to bottom, we compare the index log-returns ($r$), mean market correlation ($\mu$), volatility of the market index $r$ estimated using GARCH(1,1) process, risk ($\sigma_p$) corresponding to the minimum risk Markowitz portfolio of all the stocks in the market, network entropy (NE), communication efficiency (CE), average Ollivier-Ricci edge curvature (ORE), average Forman-Ricci edge curvature (FRE), $L^1$-norm and $L^2$-norm of persistence landscape and persistent entropy (PE). The vertical dashed lines and grey-shaded regions indicate the epochs of important events in the BSE market: 2004 Indian general election, Chinese market 2015-16 sell-off, COVID-19 Recession and Ukraine Russia war.}
    \label{fig:02_bse_time_series_085_SI}
\end{figure}
\begin{figure}[!ht]
    \centering
    \includegraphics[width=\textwidth,keepaspectratio]{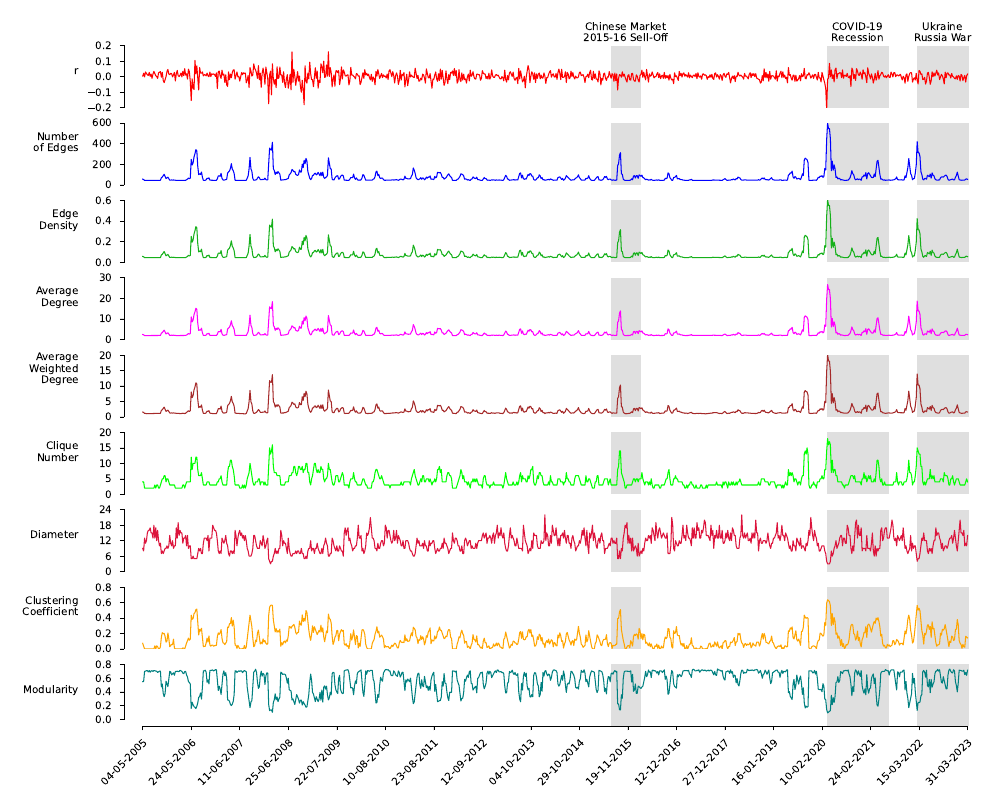}
    \caption{\textbf{Evolution of standard network measures computed on the correlation-based threshold networks $S^{\tau}(t)$ of stocks from National Stock Exchange (NSE), constructed by adding edges with correlation $C^{\tau}_{ij}(t) \geq 0.65$ to the MST, for epochs of size $\tau = 22$ days and overlapping shift $\Delta{\tau} = 5$ days.} From top to bottom, we compare the plot of index log-returns $r$ with standard network measures, namely, number of edges, edge density, average degree, average weighted degree, clique number, diameter, clustering coefficient and modularity. The grey-shaded regions indicate the epochs of important events in the NSE market: Chinese market 2015-16 sell-off, COVID-19 Recession and Ukraine Russia war.}
    \label{fig:04_nse_gt_time_series_065_SI}
\end{figure}
\begin{figure}[!ht]
    \centering
    \includegraphics[width=\textwidth,keepaspectratio]{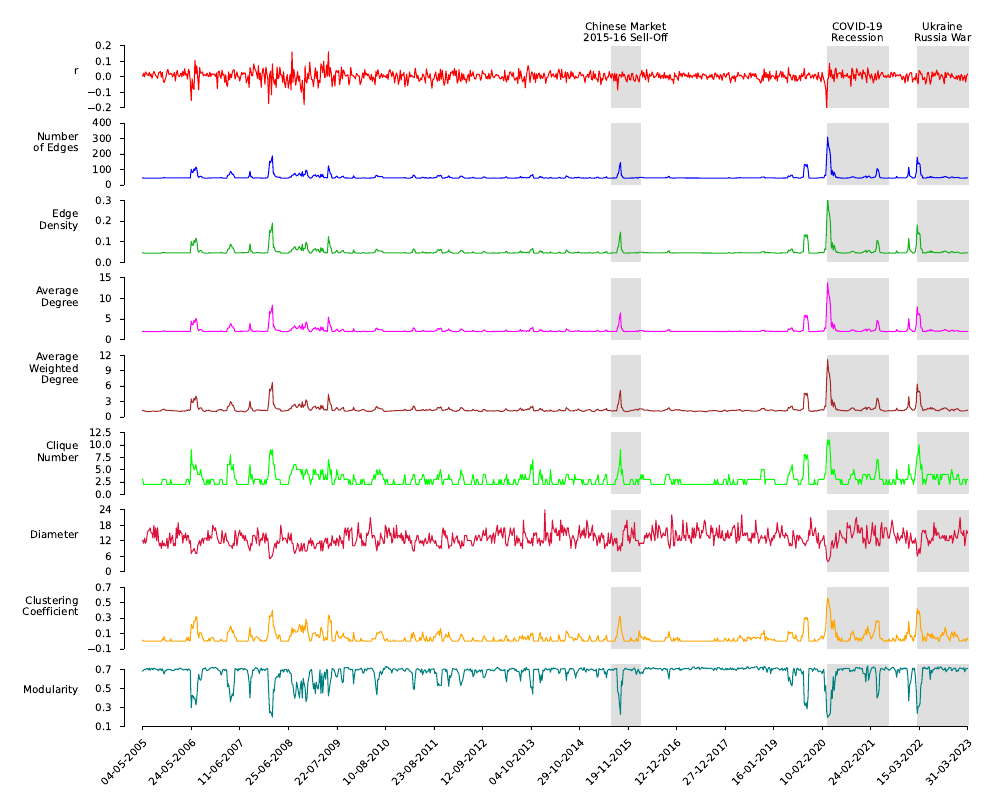}
    \caption{\textbf{Evolution of standard network measures computed on the correlation-based threshold networks $S^{\tau}(t)$ of stocks from National Stock Exchange (NSE), constructed by adding edges with correlation $C^{\tau}_{ij}(t) \geq 0.75$ to the MST, for epochs of size $\tau = 22$ days and overlapping shift $\Delta{\tau} = 5$ days.} From top to bottom, we compare the plot of index log-returns $r$ with standard network measures, namely, number of edges, edge density, average degree, average weighted degree, clique number, diameter, clustering coefficient and modularity. The grey-shaded regions indicate the epochs of important events in the NSE market: Chinese market 2015-16 sell-off, COVID-19 Recession and Ukraine Russia war.}
    \label{fig:04_nse_gt_time_series_075_SI}
\end{figure}
\begin{figure}[!ht]
    \centering
    \includegraphics[width=\textwidth,keepaspectratio]{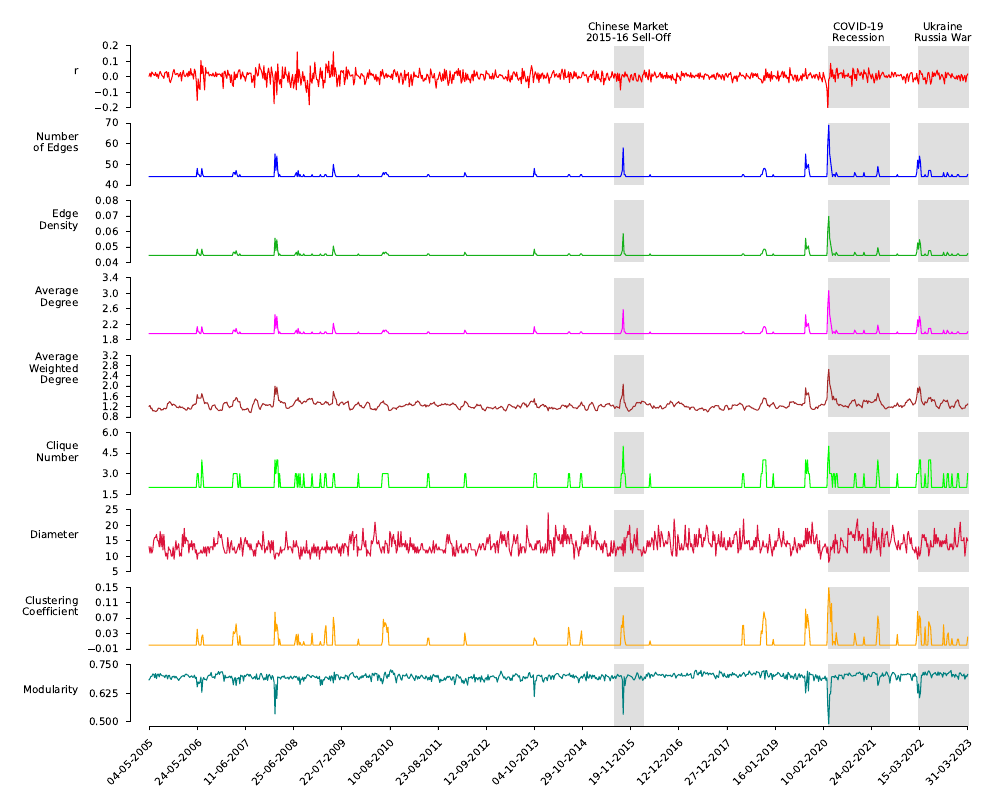}
    \caption{\textbf{Evolution of standard network measures computed on the correlation-based threshold networks $S^{\tau}(t)$ of stocks from National Stock Exchange (NSE), constructed by adding edges with correlation $C^{\tau}_{ij}(t) \geq 0.85$ to the MST, for epochs of size $\tau = 22$ days and overlapping shift $\Delta{\tau} = 5$ days.} From top to bottom, we compare the plot of index log-returns $r$ with standard network measures, namely, number of edges, edge density, average degree, average weighted degree, clique number, diameter, clustering coefficient and modularity. The grey-shaded regions indicate the epochs of important events in the NSE market: Chinese market 2015-16 sell-off, COVID-19 Recession and Ukraine Russia war.}
    \label{fig:04_nse_gt_time_series_085_SI}
\end{figure}
\begin{figure}[!ht]
    \centering
    \includegraphics[width=\textwidth,keepaspectratio]{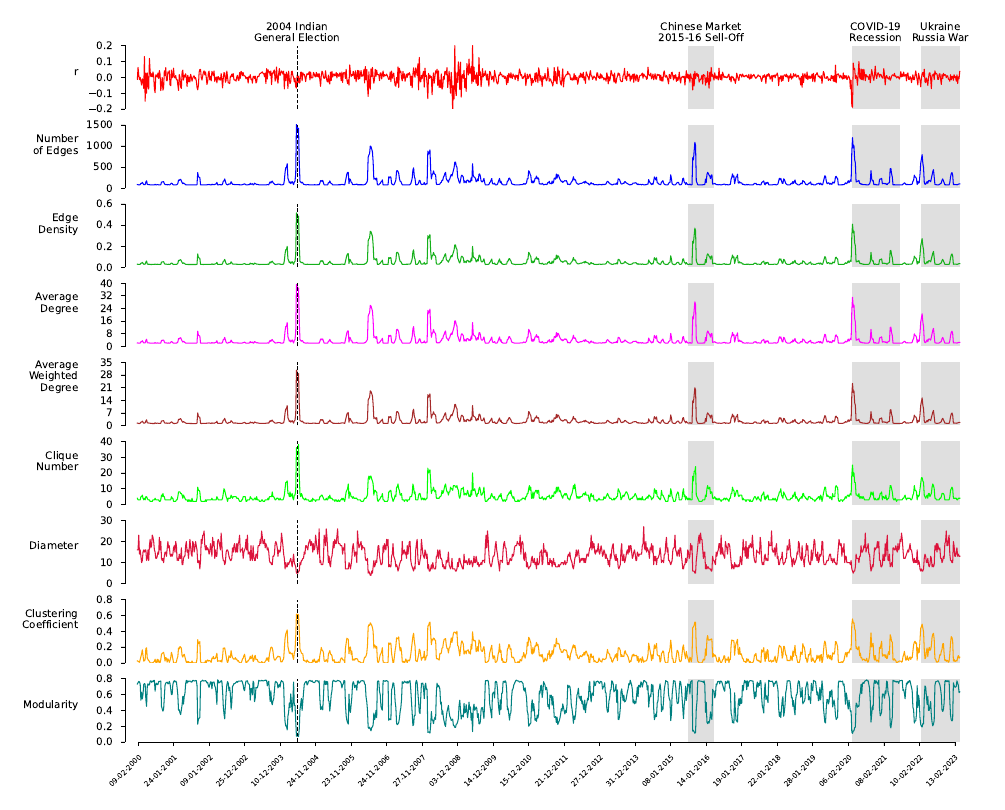}
    \caption{\textbf{Evolution of standard network measures computed on the correlation-based threshold networks $S^{\tau}(t)$ of stocks ffom Bombay Stock Exchange (BSE), constructed by adding edges with correlation $C^{\tau}_{ij}(t) \geq 0.65$ to the MST, for epochs of size $\tau = 22$ days and overlapping shift $\Delta{\tau} = 5$ days.} From top to bottom, we compare the plot of index log-returns $r$ with standard network measures, namely, number of edges, edge density, average degree, average weighted degree, clique number, diameter, clustering coefficient and modularity. The vertical dashed lines and grey-shaded regions indicate the epochs of important events in the BSE market: 2004 Indian general election, Chinese market 2015-16 sell-off, COVID-19 Recession and Ukraine Russia war.}
    \label{fig:05_bse_gt_time_series_065_SI}
\end{figure}
\begin{figure}[!ht]
    \centering
    \includegraphics[width=\textwidth,keepaspectratio]{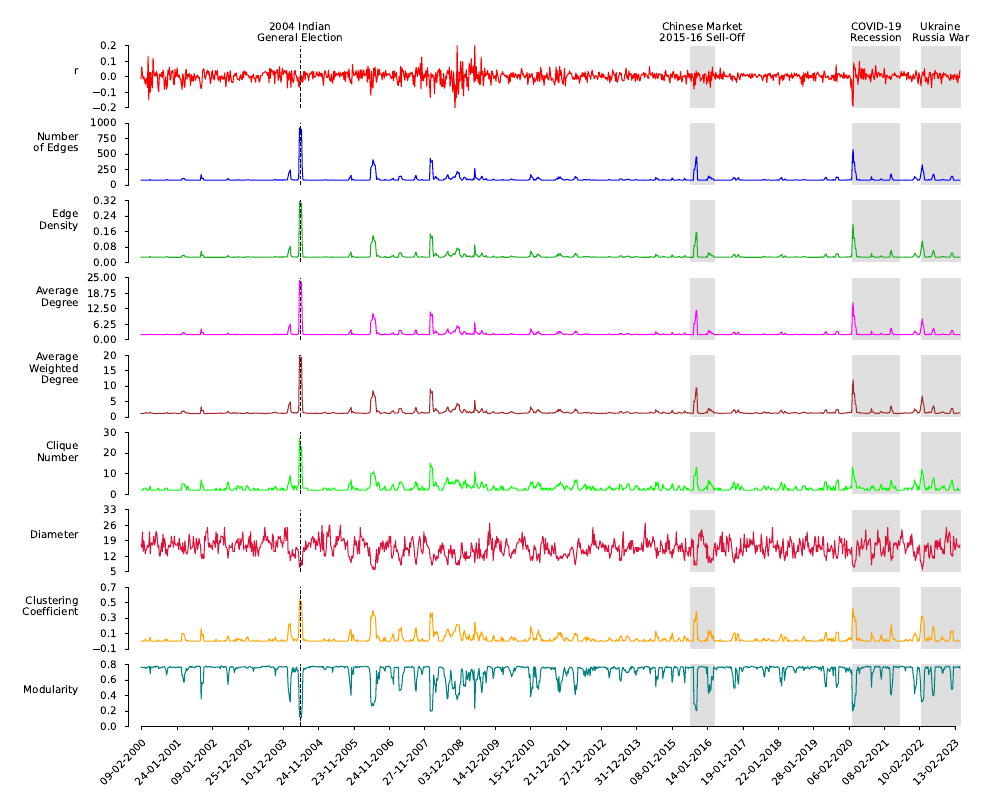}
    \caption{\textbf{Evolution of standard network measures computed on the correlation-based threshold networks $S^{\tau}(t)$ of stocks from Bombay Stock Exchange (BSE), constructed by adding edges with correlation $C^{\tau}_{ij}(t) \geq 0.75$ to the MST, for epochs of size $\tau = 22$ days and overlapping shift $\Delta{\tau} = 5$ days.} From top to bottom, we compare the plot of index log-returns $r$ with standard network measures, namely, number of edges, edge density, average degree, average weighted degree, clique number, diameter, clustering coefficient and modularity. The vertical dashed lines and grey-shaded regions indicate the epochs of important events in the BSE market: 2004 Indian general election, Chinese market 2015-16 sell-off, COVID-19 Recession and Ukraine Russia war.}
    \label{fig:05_bse_gt_time_series_075_SI}
\end{figure}
\begin{figure}[!ht]
    \centering
    \includegraphics[width=\textwidth,keepaspectratio]{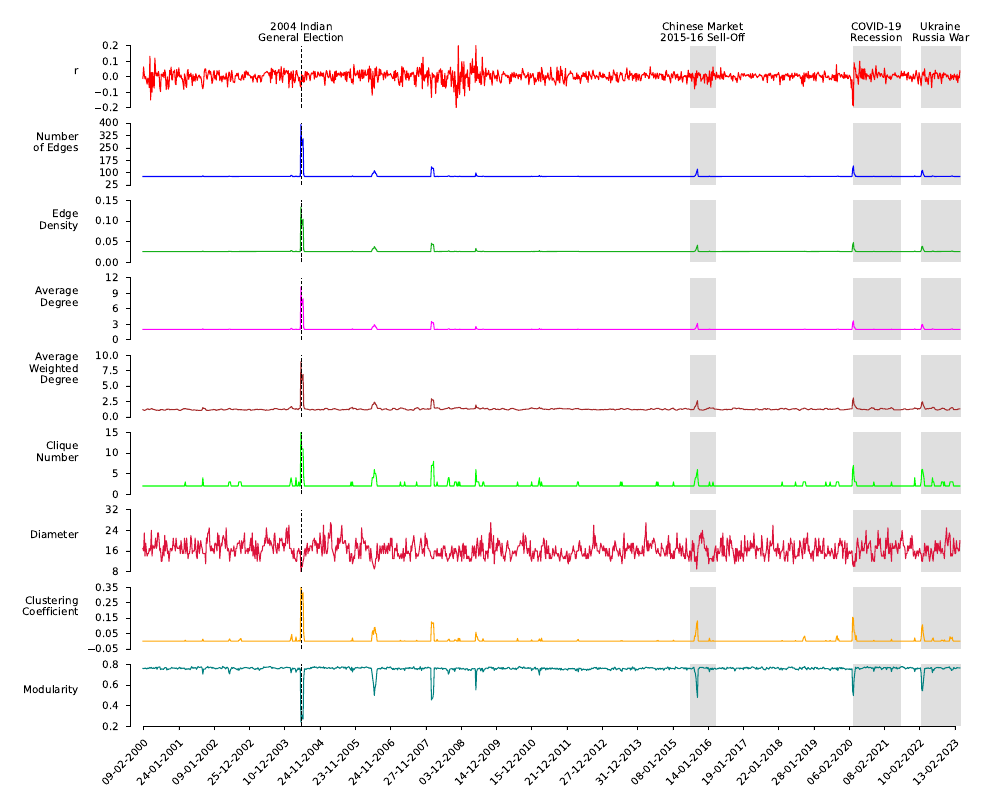}
    \caption{\textbf{Evolution of standard network measures computed on the correlation-based threshold networks $S^{\tau}(t)$ of stocks from Bombay Stock Exchange (BSE), constructed by adding edges with correlation $C^{\tau}_{ij}(t) \geq 0.85$ to the MST, for epochs of size $\tau = 22$ days and overlapping shift $\Delta{\tau} = 5$ days.} From top to bottom, we compare the plot of index log-returns $r$ with standard network measures, namely, number of edges, edge density, average degree, average weighted degree, clique number, diameter, clustering coefficient and modularity. The vertical dashed lines and grey-shaded regions indicate the epochs of important events in the BSE market: 2004 Indian general election, Chinese market 2015-16 sell-off, COVID-19 Recession and Ukraine Russia war.}
    \label{fig:05_bse_gt_time_series_085_SI}
\end{figure}
\begin{figure}[ht]
    \centering
    \includegraphics[width=\textwidth,keepaspectratio]{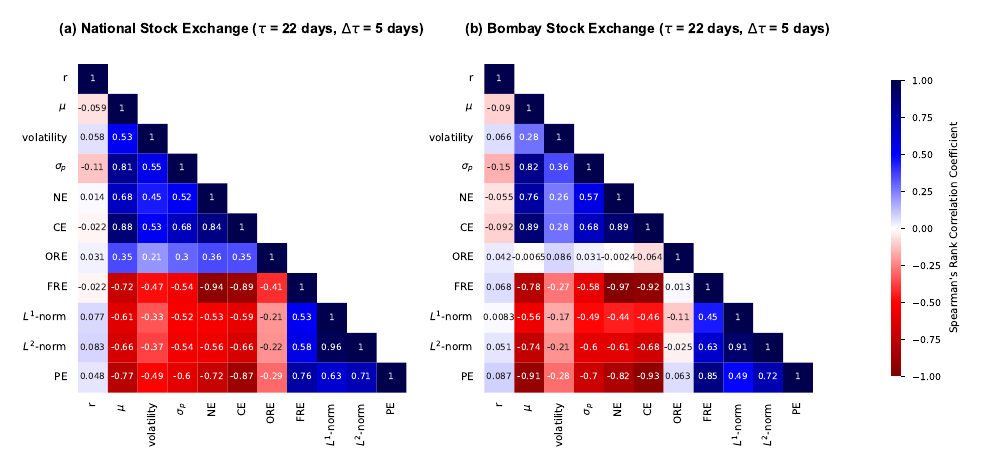}
    \caption{\textbf{Correlograms of (a) National Stock Exchange (NSE) and (b) Bombay Stock Exchange (BSE).} We show values of Spearman's rank correlation coefficient between all possible pairs of the traditional market indicators (index returns $r$, mean market correlation $\mu$, volatility, and minimum risk portfolio $\sigma_p$), standard network measures (network entropy NE, communication efficiency CE), edge-centric discrete Ricci curvatures (ORE, FRE) and persistent homology based topological measures ($L^1$-norm, $L^2$-norm, PE). The standard network measures and edge-centric discrete Ricci curvatures are computed on the correlation-based threshold networks $S^{\tau}(t)$ of stocks, constructed by adding edges with correlation $C^{\tau}_{ij}(t) \geq 0.75$ to the MST. The topological measures are computed from the persistent homology of distance matrices $\scriptstyle D^{\tau}(t) = \sqrt{2 (1 - C^{\tau}(t))}$. Each $S^{\tau}(t)$ and $D^{\tau}(t)$ correspond to the epoch size of $\tau = 22$ days with overlapping shift $\Delta{\tau} = 5$ days.}
    \label{fig:03_correlograms_075_SI}
\end{figure}
\begin{figure}[!ht]
    \centering
    \includegraphics[width=\textwidth,keepaspectratio]{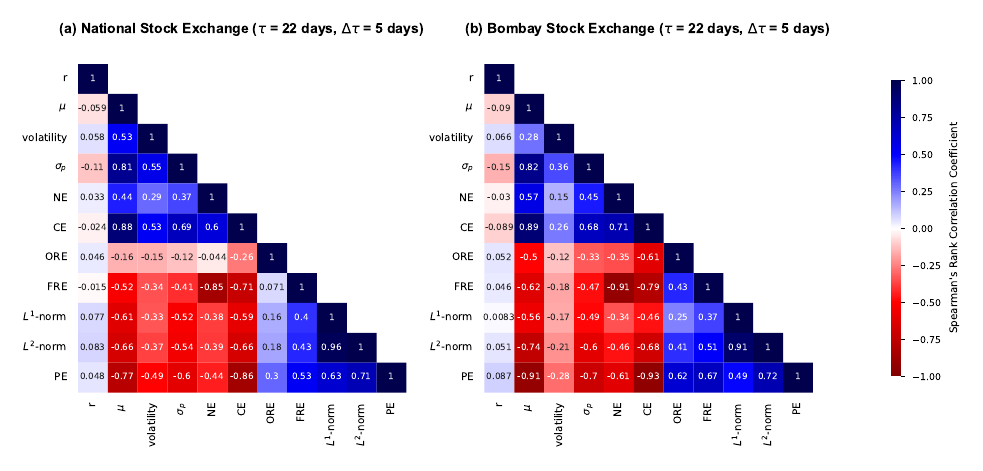}
    \caption{\textbf{Correlograms of (a) National Stock Exchange (NSE) and (b) Bombay Stock Exchange (BSE).} We show values of Spearman's rank correlation coefficient between all possible pairs of the traditional market indicators (index returns $r$, mean market correlation $\mu$, volatility, and minimum risk portfolio $\sigma_p$), standard network measures (network entropy NE, communication efficiency CE), edge-centric discrete Ricci curvatures (ORE, FRE) and persistent homology based topological measures ($L^1$-norm, $L^2$-norm, PE). The standard network measures and edge-centric discrete Ricci curvatures are computed on the correlation-based threshold networks $S^{\tau}(t)$ of stocks, constructed by adding edges with correlation $C^{\tau}_{ij}(t) \geq 0.85$ to the MST. The topological measures are computed from the persistent homology of distance matrices $\scriptstyle D^{\tau}(t) = \sqrt{2 (1 - C^{\tau}(t))}$. Each $S^{\tau}(t)$ and $D^{\tau}(t)$ correspond to the epoch size of $\tau = 22$ days with overlapping shift $\Delta{\tau} = 5$ days.}
    \label{fig:03_correlograms_085_SI}
\end{figure}
\clearpage

\begin{table}[!htbp]
\centering
\caption{\textbf{List of all stocks from National Stock Exchange (NSE) considered in this study.} The first column gives the serial number, the second column gives the symbols, the third column gives the full name of the stock, and the fourth column specifies the sector of the stock.}
\label{tab:NSE_Stocks}
\resizebox{\textwidth}{!}{%
\begin{tabular}{p{1cm}p{2.5cm}p{8cm}p{3.5cm}}
\hline
S. No. & Code       & Company Name                           & Sector                     \\ \hline
1     & TITAN      & Titan Company Ltd.                     & Consumer Discretionary     \\
2     & ASIANPAINT & Asian Paints Ltd.                      & Consumer Discretionary     \\
3     & M\&M       & Mahindra \& Mahindra Ltd.              & Consumer Discretionary     \\
4     & MARUTI     & Maruti Suzuki India Ltd.               & Consumer Discretionary     \\
5     & HEROMOTOCO & Hero Motocorp Ltd.                     & Consumer Discretionary     \\
6     & BAJAJ-AUTO & Bajaj Auto Ltd.                        & Consumer Discretionary     \\
7     & EICHERMOT  & Eicher Motors Ltd.                     & Consumer Discretionary     \\
8     & HINDUNILVR & Hindustan Unilever Ltd.                & Consumer Goods \\
9     & ITC        & ITC Ltd.                               & Consumer Goods \\
10    & BRITANNIA  & Britannia Industries Ltd.              & Consumer Goods \\
11    & ADANIENT   & Adani Enterprises Ltd.                 & Commodities                \\
12    & TATASTEEL  & Tata Steel Ltd.                        & Commodities                \\
13    & HINDALCO   & Hindalco Industries Ltd.               & Commodities                \\
14    & VEDL       & Vedanta Ltd.                        & Commodities                \\
15    & JSWSTEEL   & JSW Steel Ltd.                         & Commodities                \\
16    & UPL        & UPL Ltd.                               & Commodities                \\
17    & ULTRACEMCO & Ultratech Cement Ltd.                  & Commodities                \\
18    & SHREECEM   & Shree Cement Ltd.                      & Commodities                \\
19    & RELIANCE   & Reliance Industries Ltd.               & Energy                     \\
20    & ONGC       & Oil And Natural Gas Corporation Ltd    & Energy                     \\
21    & IOC        & Indian Oil Corporation Ltd.            & Energy                     \\
22    & BPCL       & Bharat Petroleum Corporation Ltd.      & Energy                     \\
23    & GAIL       & GAIL (India) Ltd.                      & Energy                     \\
24    & HINDPETRO  & Hindustan Petroleum Corporation Ltd.   & Energy                     \\
25    & HDFCBANK   & HDFC Bank Ltd                          & Financial Services         \\
26    & ICICIBANK  & ICICI Bank Ltd.                        & Financial Services         \\
27    & KOTAKBANK  & Kotak Mahindra Bank Ltd.               & Financial Services         \\
28    & AXISBANK   & Axis Bank Ltd.                         & Financial Services         \\
29    & SBIN       & State Bank Of India                    & Financial Services         \\
30    & INDUSINDBK & Indusind Bank Ltd.                     & Financial Services         \\
31    & HDFC       & Housing Development Finance Corp. Ltd. & Financial Services         \\
32    & BAJFINANCE & Bajaj Finance Ltd.                  & Financial Services         \\
33    & DRREDDY    & Dr.Reddy's Laboratories Ltd.           & Healthcare                 \\
34    & CIPLA      & Cipla Ltd.                             & Healthcare                 \\
35    & SUNPHARMA  & Sun Pharmaceutical Industries Ltd.     & Healthcare                 \\
36    & DIVISLAB   & Divi's Laboratories Ltd.               & Healthcare                 \\
37    & LUPIN      & Lupin Ltd.                             & Healthcare                 \\
38    & APOLLOHOSP & Apollo Hospitals Enterprise Ltd.       & Healthcare                 \\
39    & LT         & Larsen \& Toubro Ltd.                  & Industrials                \\
40    & TCS        & Tata Consultancy Services Ltd.         & Information Technology     \\
41    & INFY       & Infosys Ltd.                           & Information Technology     \\
42    & HCLTECH    & HCL Technologies Ltd.                  & Information Technology     \\
43    & WIPRO      & Wipro Ltd.                             & Information Technology     \\
44    & NTPC       & NTPC Ltd.                              & Utilities                  \\
45    & TATAPOWER  & Tata Power Co. Ltd.                    & Utilities                  \\ \hline
\end{tabular}%
}
\end{table}

\begin{table}[!htbp]
\centering
\caption{\textbf{List of all stocks from Bombay Stock Exchange (BSE) considered in this study.} The first column gives the serial number, the second column gives the symbols, the third column gives the full name of the stock, and the fourth column specifies the sector of the stock.}
\label{tab:BSE_Stocks_1}
\resizebox{\textwidth}{!}{%
\begin{tabular}{p{0.8cm}p{2.5cm}p{8cm}p{3.5cm}}
\hline
S. No. & Code       & Company Name                                          & Sector                     \\ \hline
1     & BOMDYEING  & Bombay Dyeing \& Mfg. Co. Ltd.                          & Consumer Discretionary     \\
2     & DEEPAKSP   & Deepak Spinners Ltd.                                  & Consumer Discretionary     \\
3     & EICHERMOT  & Eicher Motors Ltd.                                    & Consumer Discretionary     \\
4     & EIHOTEL    & EIH Ltd.                                              & Consumer Discretionary     \\
5     & HBLPOWER   & HBL Power Systems Ltd.                                & Consumer Discretionary     \\
6     & HIL        & HIL Ltd.                                              & Consumer Discretionary     \\
7     & RAYMOND    & Raymond Ltd.                                          & Consumer Discretionary     \\
8     & TATAMOTORS & Tata Motors Ltd.                                      & Consumer Discretionary     \\
9     & TITAN      & Titan Company Ltd.                                    & Consumer Discretionary     \\
10    & BRITANNIA  & Britannia Industries Ltd.                             & Consumer Goods \\
11    & COLPAL     & Colgate-Palmolive (India) Ltd.                        & Consumer Goods \\
12    & COMFINTE   & Comfort Intech Ltd.                                   & Consumer Goods \\
13    & HINDUNILVR & Hindustan Unilever Ltd.                               & Consumer Goods \\
14    & ITC        & ITC Ltd.                                              & Consumer Goods \\
15    & KSE        & KSE Ltd.                                              & Consumer Goods \\
16    & MARICO     & Marico Ltd.                                           & Consumer Goods \\
17    & TASTYBIT   & Tasty Bite Eatables Ltd.                              & Consumer Goods \\
18    & AMBUJACEM  & Ambuja Cements Ltd.                                   & Commodities                \\
19    & ASHAPURMIN & Ashapura Minechem Ltd.                                & Commodities                \\
20    & DEEPAKFERT & Deepak Fertilisers \& Petrochemicals Corporation Ltd. & Commodities                \\
21    & GSFC       & Gujarat State Fertilizers \& Chemicals Ltd.           & Commodities                \\
22    & HEIDELBERG & Heidelbergcement India Ltd.                           & Commodities                \\
23    & HINDALCO   & Hindalco Industries Ltd.                              & Commodities                \\
24    & HINDZINC   & Hindustan Zinc Ltd.                                   & Commodities                \\
25    & INDIACEM   & India Cements Ltd.                                    & Commodities                \\
26    & KESORAMIND & Kesoram Industries Ltd.                               & Commodities                \\
27    & LINDEINDIA & Linde India Ltd.                                   & Commodities                \\
28    & MANGCHEFER & Mangalore Chemicals \& Fertilizers Ltd.               & Commodities                \\
29    & NATIONALUM & National Aluminium Co.Ltd.                            & Commodities                \\
30    & SAIL       & Steel Authority Of India Ltd.                         & Commodities                \\
31    & SHREDIGCEM & Shree Digvijay Cement Co.Ltd.                         & Commodities                \\
32    & VEDL       & Vedanta Ltd.                                       & Commodities                \\
33    & VINATIORGA & Vinati Organics Ltd.                                  & Commodities                \\
34    & BPCL       & Bharat Petroleum Corporation Ltd.                     & Energy                     \\
35    & GAIL       & Gail (India) Ltd.                                     & Energy                     \\
36    & HINDOILEXP & Hindustan Oil Exploration Co.Ltd.                     & Energy                     \\
37    & HINDPETRO  & Hindustan Petroleum Corporation Ltd.                  & Energy                     \\
38    & MRPL       & Mangalore Refinery \& Petrochemicals Ltd.             & Energy                     \\
39    & ONGC       & Oil And Natural Gas Corporation Ltd.                   & Energy                     \\
40    & RELIANCE   & Reliance Industries Ltd.                              & Energy                     \\
41    & AXISBANK   & Axis Bank Ltd.                                        & Financial Services         \\
42    & FEDERALBNK & Federal Bank Ltd.                                     & Financial Services         \\
43    & HDFCBANK   & HDFC Bank Ltd.                                         & Financial Services         \\
44    & IDBI       & IDBI Bank Ltd.                                        & Financial Services         \\
45    & KOTAKBANK  & Kotak Mahindra Bank Ltd.                              & Financial Services         \\ \hline
\end{tabular}%
}
\end{table}
\begin{table}[]
\centering
\resizebox{\textwidth}{!}{%
\begin{tabular}{p{0.8cm}p{2.5cm}p{8cm}p{3.5cm}}
\hline
S. No. & Code       & Company Name                                          & Sector                     \\ \hline
46    & MFSL       & Max Financial Services Ltd.                          & Financial Services         \\
47    & PEL        & Piramal Enterprises Ltd.                              & Financial Services         \\
48    & SBIN       & State Bank Of India                                   & Financial Services         \\
49    & SOUTHBANK  & South Indian Bank Ltd.                                & Financial Services         \\
50    & TATAINVEST & Tata Investment Corporation Ltd.                      & Financial Services         \\ 
51    & AUROPHARMA & Aurobindo Pharma Ltd.                                 & Healthcare                 \\
52    & CIPLA      & Cipla Ltd.                                            & Healthcare                 \\
53    & DRREDDY    & Dr.Reddy's Laboratories Ltd.                          & Healthcare                 \\
54    & INDRAMEDCO & Indraprastha Medical Corp.Ltd.                        & Healthcare                 \\
55    & LUPIN      & Lupin Ltd.                                            & Healthcare                 \\
56    & LYKALABS   & Lyka Labs Ltd.                                        & Healthcare                 \\
57    & MOREPENLAB & Morepen Laboratories Ltd.                             & Healthcare                 \\
58    & SUNPHARMA  & Sun Pharmaceutical Industries Ltd.                    & Healthcare                 \\
59    & UNICHEMLAB & Unichem Laboratories Ltd.                             & Healthcare                 \\
60    & BHARATFORG & Bharat Forge Ltd.                                     & Industrials                \\
61    & BHEL       & Bharat Heavy Electricals Ltd.                         & Industrials                \\
62    & ENGINERSIN & Engineers India Ltd.                                  & Industrials                \\
63    & ESCORTS    & Escorts Kubota Ltd.                                    & Industrials                \\
64    & HCC        & Hindustan Construction Co. Ltd.                        & Industrials                \\
65    & HITECHCORP & Hitech Corporation Ltd.                             & Industrials                \\
66    & INDLMETER  & Imp Powers Ltd.                                       & Industrials                \\
67    & KENNAMET   & Kennametal India Ltd.                              & Industrials                \\
68    & LT         & Larsen \& Toubro Ltd.                                 & Industrials                \\
69    & SIEMENS    & Siemens Ltd.                                          & Industrials                \\
70    & INFY       & Infosys Ltd.                                          & Information Technology     \\
71    & SONATSOFTW & Sonata Software Ltd.                                  & Information Technology     \\
72    & WIPRO      & Wipro Ltd.                                            & Information Technology     \\
73    & SPICEJET   & Spicejet Ltd.                                         & Services                   \\
74    & MTNL       & Mahanagar Telephone Nigam Ltd.                        & Telecommunication          \\
75    & VINDHYATEL & Vindhya Telelinks Ltd.                                & Telecommunication          \\
76    & NLCINDIA   & NLC India Ltd.                                      & Utilities                  \\
77    & RELINFRA   & Reliance Infrastructure Ltd.                          & Utilities                  \\ \hline
\end{tabular}%
}
\end{table}

\begin{table}[ht]
\centering
\caption{\textbf{The categorization of standard standard network measures employed in this study.} The measures used to analyze the correlation-based threshold networks of stocks from NSE and BSE are divided into two groups of unweighted and weighted measures. While computing each measure in the weighted category, we provide a corresponding natural weight, strength or distance, that is to be utilized in the corresponding computation.}
\label{tab:network_measures}
\begin{tabular}{clcc}
\hline
S. No. & Network measure            & Type of measure & Choice of weight \\ \hline
1       & Edge count                 & Unweighted      & -                \\
2       & Edge density               & Unweighted      & -                \\
3       & Average degree             & Unweighted      & -                \\
4       & Clique number              & Unweighted      & -                \\
5       & Network entropy            & Unweighted      & -                \\
6       & Average weighted degree           & Weighted        & Strength         \\
7       & Global assortativity       & Weighted        & Strength         \\
8       & Average clustering coefficient     & Weighted        & Strength         \\
9       & Modularity                 & Weighted        & Strength         \\
10      & Diameter                   & Weighted        & Distance         \\
11      & Global reaching centrality & Weighted        & Distance         \\
12      & Communication efficiency   & Weighted        & Distance         \\
13      & Ollivier-Ricci curvature   & Weighted        & Distance         \\
14      & Forman-Ricci curvature     & Weighted        & Distance         \\ \hline
\end{tabular}
\end{table}

\end{document}